\documentclass[twocolumn,prb,amsmath,amssymb,floatfix,superscriptaddress]{revtex4-2} 
\usepackage{amsmath}
\usepackage{amssymb}
\usepackage{graphicx}
\usepackage{bm}
\usepackage{color}
\usepackage{hyperref}
\usepackage{graphicx}
\usepackage{upgreek}
\usepackage{scalerel}
\usepackage{multirow}
 
\usepackage{pgffor}
\usepackage{pdfpages}
\usepackage{mathdots}
\usepackage{float}
\usepackage{silence}
\usepackage{physics} 
\usepackage{lscape}   
\usepackage{color, colortbl}
\usepackage[table]{xcolor}
\usepackage{comment}  
\makeatletter
\AtBeginDocument{\let\LS@rot\@undefined}
\makeatother

\begin{document}

\title{Controllable odd-frequency Cooper pairs in multi-superconductor Josephson junctions}
\author{Jorge Cayao}
\affiliation{Department of Physics and Astronomy, Uppsala University, Box 516, S-751 20 Uppsala, Sweden}

\author{Pablo Burset}
\affiliation{Department of Theoretical Condensed Matter Physics, Condensed Matter Physics Center (IFIMAC) and Instituto Nicol\'as Cabrera, Universidad Aut\'onoma de Madrid, 28049 Madrid, Spain}

\author{Yukio Tanaka}
\affiliation{Department of Applied Physics, Nagoya University, Nagoya 464-8603, Japan}
\affiliation{Research Center for Crystalline Materials Engineering, Nagoya University, Nagoya 464-8603, Japan}

\date{\today}
\begin{abstract}
We consider  Josephson junctions formed by multiple superconductors with distinct phases and explore the formation of nonlocal or inter-superconductor pair correlations. We find that the multiple superconductor nature offers an additional degree of freedom that broadens the classification of pair symmetries, enabling nonlocal even- and odd-frequency pairings that can be highly controlled by the superconducting phases and the  energy of the superconductors. Specially, when the phase difference between two superconductors is $\pi$, their associated nonlocal odd-frequency pairing is the only type of inter-superconductor pair correlations.  Finally, we show that these nonlocal odd-frequency Cooper pairs dominate the nonlocal conductance  via crossed Andreev reflections, which constitutes a direct evidence of odd-frequency pairing.  
\end{abstract}
\maketitle
\section{Introduction}
Superconductivity is caused by electrons binding together into  Cooper pairs
 below a critical temperature and  has attracted great interest due to its  properties for   quantum technologies \cite{Ac_n_2018}. The applications of superconductors are thus intimately linked to the Cooper pairs, specially  to the symmetries or their wavefunction or pair amplitude.  Due to the fermionic nature of electrons, the pair amplitude is antisymmetric under the exchange of all the quantum numbers describing the paired electron states plus the exchange of their relative time coordinates. Of particular interest is that the antisymmetry  enables the formation of odd-frequency Cooper pairs, where the pair amplitude is odd in the relative time, or frequency $\omega$, of the paired electrons \cite{bere74,PhysRevLett.66.1533,PhysRevB.45.13125,PhysRevB.46.8393,PhysRevB.52.1271,coleman1997reflections}. As a result, the odd-$\omega$ Cooper pairs characterize a unique type of superconducting pairing that is intrinsically dynamic \cite{RevModPhys.77.1321,Nagaosa12,Balatsky2017,cayao2019odd,triola2020role,tanaka2021theory}.

Odd-$\omega$ Cooper pairs have been studied as bulk and induced effects in several systems \cite{RevModPhys.77.1321,Nagaosa12,Balatsky2017,cayao2019odd,triola2020role,tanaka2021theory}, such as in superconducting heterostructures \cite{PhysRevLett.86.4096,Kadigrobov01,PhysRevLett.98.037003,PhysRevB.75.134510,PhysRevLett.98.107002,PhysRevLett.99.037005,PhysRevB.76.054522,Eschrig2007,PhysRevB.78.012508,PhysRevLett.107.087001,PhysRevB.87.104513,PhysRevB.87.220506,PhysRevB.91.054518,PhysRevB.91.144514,PhysRevB.92.100507,PhysRevB.92.205424,PhysRevB.92.121404,PhysRevB.95.174516,PhysRevB.96.155426,balatsky2018oddfrequency,PhysRevB.94.014504,PhysRevB.95.224502,PhysRevB.97.075408,PhysRevB.97.134523,PhysRevB.98.075425,PhysRevB.98.134508,PhysRevB.98.161408,PhysRevB.99.184501,PhysRevB.100.024511,PhysRevB.100.115433,PhysRevB.100.144511,PhysRevB.101.024509,PhysRevB.101.064514,PhysRevB.101.094505,PhysRevB.101.094506,PhysRevResearch.2.022019,PhysRevResearch.2.033229,PhysRevResearch.2.043193,PhysRevResearch.2.043388,PhysRevB.101.195303,PhysRevB.102.140505,PhysRevB.103.024501,PhysRevB.103.184509,PhysRevB.103.245138,PhysRevB.104.054519,PhysRevB.104.094503,PhysRevB.104.144513,PhysRevB.104.165125,PhysRevResearch.3.043148,PhysRevB.105.214512,PhysRevB.105.224506,PhysRevResearch.5.L012022,PhysRevB.107.054501,PhysRevB.107.064504,PhysRevB.108.094503,tanaka2024theory}, multiband superconductors 
\cite{PhysRevB.88.104514,v2013theory,PhysRevB.90.220501,PhysRevB.92.094517,PhysRevB.92.224508,PhysRevB.93.201402,10.1093/ptep/ptw094,PhysRevLett.119.087001,asano2018dirty,PhysRevB.101.180512,PhysRevB.101.214507,PhysRevB.104.054507,PhysRevResearch.3.033255}, time-periodic superconductors \cite{PhysRevB.94.094518,PhysRevB.103.104505,kuhn2023floquet}, and non-Hermitian superconductors \cite{PhysRevB.105.094502}. There also exist experiments supporting the realization of  induced triplet odd-$\omega$ pairs in hybrid systems between superconductors with magnetic materials \cite{
keizer2006spin,PhysRevLett.96.157002,robinson2010controlled,PhysRevB.82.060505,PhysRevB.82.100501,PhysRevLett.104.137002,anwar2011inducing,PhysRevLett.108.127002,anwar2012long,
PhysRevX.5.041021,di2015signature,PhysRevLett.125.026802,PhysRevLett.125.117003,chiu2021observation,chiu2023tuning}. All these studies show that,  to induce odd-$\omega$  pairs, the symmetries linked to the quantum numbers of the paired electrons must break \cite{triola2020role}. While this condition guarantees the formation of odd-$\omega$   pairs, it does not restrict the appearance of even-$\omega$ pairs \cite{PhysRevB.104.054507,PhysRevB.106.024511,PhysRevLett.129.247001}  which then mask the   odd-$\omega$ signatures. Yet another issue is that controlling odd-$\omega$ pairs,  despite the efforts \cite{PhysRevLett.109.057005,PhysRevB.91.060501,banerjee2014reversible,PhysRevB.90.134514,PhysRevLett.116.077001,glick2018phase}, is still challenging without magnetic materials.

\begin{figure}[!t]
\centering
	\includegraphics[width=0.8\columnwidth]{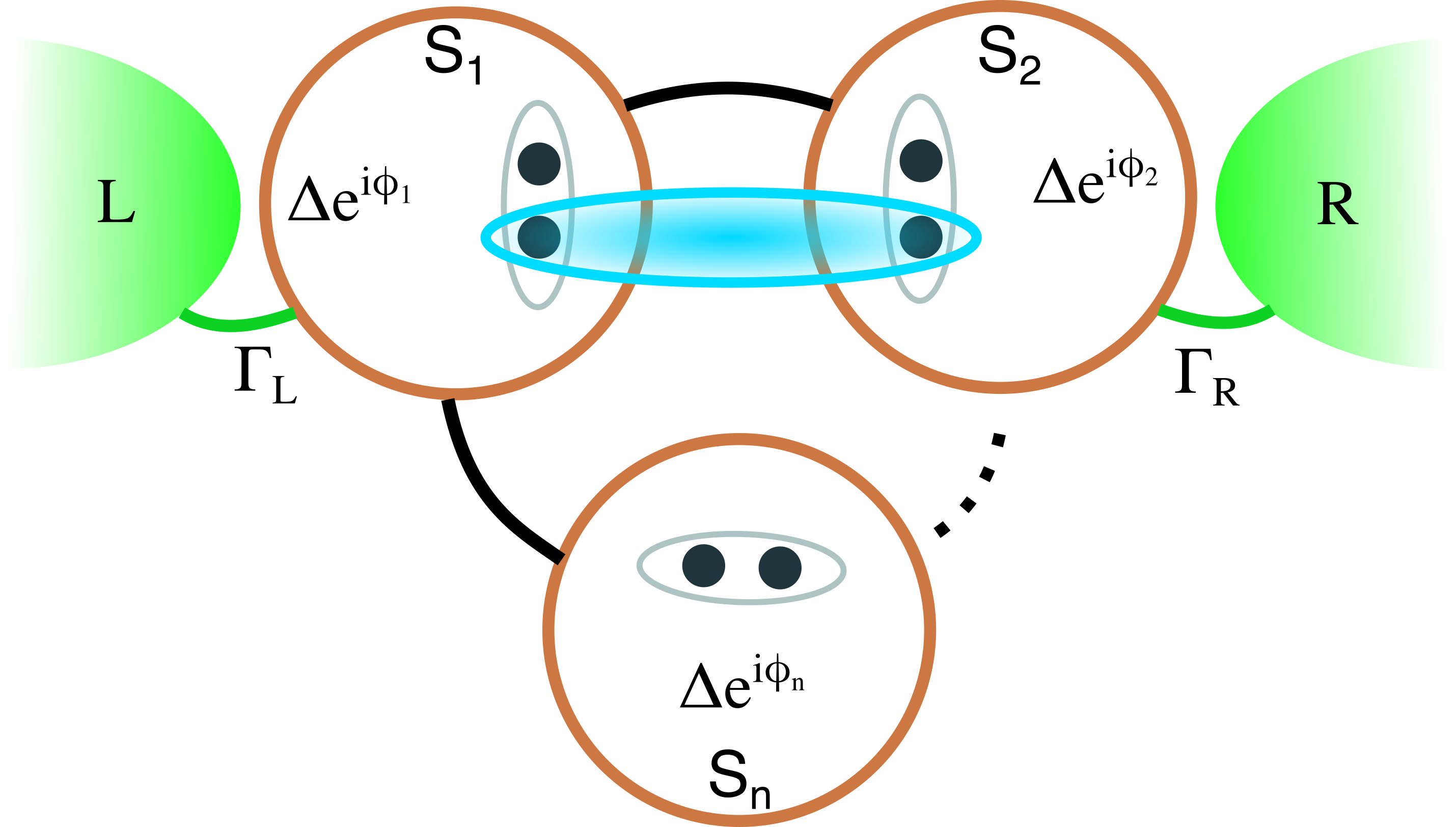}\\
 	\caption{JJs formed by coupling superconductors $S_{i}$ with distinct phases $\phi_{i}$, and same induced pair potential $\Delta$. In each $S_{i}$  local pairs are depicted in gray ellipses containing two electrons (black filled circles),   referred to as intra superconductor (local) pairs.  Due to the tunneling between superconductors, inter-superconductor (nonlocal)  pair correlations emerge (cyan) which can be controlled by $\phi_{i}$.   Normal leads (green) are attached to two $S_{i}$ for exploring nonlocal  transport and detecting inter-superconductor Cooper pairs.}
\label{Fig0} 
\end{figure}

In this work we demonstrate the generation, control, and  direct detection of spin-singlet odd-$\omega$ Cooper pairs in Josephson junctions (JJs) formed by multiple superconductors [Fig.\,\ref{Fig0}]. In particular, we exploit the degree of freedom offered by the multi superconductor nature of the setup and find that inter-superconductor even- and odd-$\omega$ Cooper pairs naturally arise and can be controlled by the superconducting phases and onsite energies of the superconductors. Interestingly, for a JJ with two superconductors, the even-$\omega$ amplitude vanishes either when  the superconducting phase difference is $\pi$ or at zero onsite energy, leaving  only odd-$\omega$ pairing. This behaviour remains when the number of superconductors increases but only at weak couplings between superconductors. Furthermore, we discover that   crossed Andreev reflections (CARs) directly probe odd-$\omega$ Cooper pairs and can be controlled by the superconducting phases.  Our work thus  puts forward multi-superconductor JJs as a powerful and entirely different route for odd-$\omega$ Cooper pairs.

The remainder of this article is organized as follows. In Sec.~\ref{sectionII}, we introduce the multi-superconductor JJs studied in this work, while in Sec.~\ref{sectionIII} we show how to obtain the emerging pair amplitudes. In Sec.\,\ref{sectionIV} we present the obtained even- and odd-$\omega$ pair amplitudes and discuss their tunability by the superconducting phases. In Sec.\,\ref{sectionV} we demonstrate how the nonlocal odd-$\omega$ pair amplitude is detected via CAR processes. Finally, in Sec.~\ref{sectionVI} we present our conclusions. 

\section{Multi-superconductor JJs}
\label{sectionII}
We consider  JJs as shown in Fig.\,\ref{Fig0},   where $n$ conventional spin-singlet $s$-wave superconductors are coupled directly. For the sake of simplicity, we model these JJs by only considering the contact regions, with a Hamiltonian given by  
\begin{equation}
\label{JJEqs}
\begin{split}
H_{\rm nJJ}&=\sum_{j=1}^{n}[\epsilon_{j}c^{\dagger}_{j\sigma}c_{j\sigma}+\Delta{\rm e}^{i\phi_{j}}c_{j\sigma}^{\dagger}c_{j\bar{\sigma}}^{\dagger}+ h.\,c.]+H_{\rm T} \,,\\
\end{split}
\end{equation}
where the first two terms describe the superconductor S$_{j}$, where  $c_{j\sigma}$ ($c^{\dagger}_{j\sigma}$) annihilates (creates) an electronic state with spin $\sigma$ at site $j$  with onsite energy $\epsilon_{j}$,  phase $\phi_{j}$, and induced pair potential   $\Delta$  from a parent spin-singlet $s$-wave superconductor with order parameter $\Delta_{\rm sc}$.   Moreover,  $H_{\rm T}=t_{0}\sum_{j=1}^{n}c_{j\sigma}^{\dagger}c_{j+1\sigma}+{\rm h.c.}$ represents the coupling between superconductors with equal strength $t_{0}$ and  $c_{n+1}=c_{1}$.
  Away from the bulk gap edges, $\Delta$ is determined as $\Delta=\tau^{2}/\Delta_{\rm sc}$ \cite{PhysRevB.54.7366,PhysRevB.64.104502,PhysRevB.63.134515,bauer2007spectral} where $\tau$ is the coupling between S$_{j}$ and the bulk superconductor. Below we choose $\tau=0.7$ and $\Delta=0.5$ such that  $\Delta_{\rm sc}=1$ is larger than the induced gap and fix it as our energy unit.  We also drop the spin index for simplicity but keep in mind that the superconductors in Eq.\,(\ref{JJEqs}) are spin-singlet.  Despite the simplicity of our model, it captures the main effects we aim to explore in this work, namely, the multi superconductor nature and the distinct superconducting phases. Systems involving multiple JJs have been studied before but in the context of topological phases \cite{van2014single,PhysRevB.92.155437,riwar2016multi,strambini2016omega,amundsen2017analytically,PhysRevB.97.035443,PhysRevB.100.014521,PhysRevB.101.174506,PhysRevX.10.031051,PhysRevResearch.1.033004,PhysRevResearch.3.013288,peralta2022multi}. Here, we expand the playground of these multi-superconductor JJs for realizing controllable  odd-$\omega$ Cooper pairs.

\section{Superconducting pair amplitudes}
\label{sectionIII}
We are interested in  inter-superconductor pair correlations which we also refer to as nonlocal pair correlations as they reside between  superconductors. Pair correlations  are described by the anomalous Green's function ${\mathcal F}_{nm}(1,1')=\langle \mathcal{T}c_{n}(1)c_{m}(1')\rangle$ where $\mathcal{T}$ is the time ordering operator, ${c_{n}}$ annihilates an electronic state with quantum numbers $n$ at time and position $1=(x_{1},t_{1})$ \cite{mahan2013many,zagoskin}. The fermionic nature of electrons dictates  the antisymmetry condition ${\mathcal F}_{nm}(1,1')=-{\mathcal F}_{mn}(1',1)$, which enables the classification of superconducting pair correlations based on all the quantum numbers, including time and space coordinates \cite{RevModPhys.77.1321,Nagaosa12,Balatsky2017,cayao2019odd,triola2020role,tanaka2021theory}. Thus, this condition enables  even- and odd-$\omega$ pair correlations when $F_{nm}(\omega)=\pm F_{nm}(-\omega)$, with $F_{nm}(\omega)$ being the Fourier transform of ${\mathcal F}_{nm}(1,1')$ into frequency domain. In the case of multi-superconductor junctions, the multiple superconductor nature introduces an additional quantum number $n$, the superconductor index, that broadens the classification of pair symmetries in a similar way as the band index in multiband superconductors \cite{triola2020role}. In Table \ref{table1} we present all the  allowed pair symmetry classes that respect the antisymmetry condition in JJs with spin-singlet and spin-triplet superconductors:   four classes correspond to odd-$\omega$ pair correlations which are the four bottom classes in Table \ref{table1}, see Supplementary Material \cite{SM} for  details.  It is evident that the superconductor index (sup. index) plays a crucial role for broadening the allowed pair symmetries.

\begin{table}[t!]
\centering
\begin{tabular}{ |c||c|c|c||c|  }
 \hline
 \multicolumn{5}{|c|}{Pair symmetries in multi-superconductor JJs} \\
 \hline
Frequency & Spin & Sup. index &Parity&Pair symmetry class\\
  ($\omega\leftrightarrow -\omega$)& ($\uparrow \leftrightarrow \downarrow$) & ($n\leftrightarrow m$)& ($x\leftrightarrow x'$)& (total exchange) \\
 \hline
 {\bf E}ven   & {\bf S}inglet    &{\bf E}ven&    {\bf E}ven & ESEE\\
 {\bf E}ven&   {\bf S}inglet  & {\bf O}dd   &{\bf O}dd&ESOO\\
 {\bf E}ven &{\bf T}riplet &  {\bf E}ven&  {\bf O}dd&ETEO\\
 {\bf E}ven    &{\bf T}riplet & {\bf O}dd&  {\bf E}ven&ETOE\\
 {\bf O}dd&  {\bf S}inglet  & {\bf E}ven&{\bf O}dd&OSEO\\
{\bf O}dd& {\bf S}inglet  & {\bf O}dd   & {\bf E}ven&OSOE\\
{\bf O}dd& {\bf T}riplet  & {\bf E}ven& {\bf E}ven&OTEE\\
{\bf O}dd& {\bf T}riplet  & {\bf O}dd&{\bf O}dd&OTOO\\
 \hline
\end{tabular}
\caption{Allowed superconducting pair symmetries in multi-superconductor JJs under the presence of spin-mixing fields. The classes ESEE and OSOE correspond to the pair correlations reported in this work.}
\label{table1}
\end{table}

In the JJs with spin-singlet $s$-wave superconductors considered here, the symmetric and antisymmetric combination  $F_{nm}^{+(-)}=(F_{nm}\pm F_{mn})/2$ become even- and odd-$\omega$ pair symmetry classes, respectively~\cite{SM}. These two pair symmetry classes correspond to the ESEE and OSOE classes indicated in orange in Table \ref{table1}. In practice, the pair correlations  $F_{nm}$ are  obtained from the electron-hole component of the Nambu Green's function, whose equation of motion in frequency space reads $[\omega-\mathcal{H}_{\rm nJJ}]G(\omega)=\mathbf{I}$, where $\mathcal{H}_{\rm nJJ} $ is the Nambu Hamiltonian of the JJ with $n$ superconductors described by Eqs.\,(\ref{JJEqs}) in the basis  $\Psi=(c_{1},c_{1}^{\dagger},c_{2},c_{2}^{\dagger},\cdots c_{n},c_{n}^{\dagger})^{\rm T}$.


\section{Inter-superconductor pair amplitudes in JJs}
\label{sectionIV}
To begin, we focus on the   pair correlations in a JJ with two superconductors coupled directly. This system is modelled by $H_{\rm 2JJ}$ with $n=2$ in Eq.\,(\ref{JJEqs}). As described in the previous section,  the pair correlations are   obtained from electron-hole components of the Green's function  associated to the Nambu Hamiltonian in the basis   $\Psi=(c_{1},c_{1}^{\dagger},c_{2},c_{2}^{\dagger})^{\rm T}$.  Without loss of generality, we assume a phase difference  $\phi_{2}-\phi_{1}=\phi$. Then, considering $\epsilon_{1,2}\equiv\epsilon$, the symmetric and antisymmetric pair amplitudes in superconductor index are given by   \cite{SM}
\begin{equation}
\label{LRLR}
\begin{split}
F_{\rm 12}^{+}(\omega)&=\frac{2 \epsilon \Delta  t_{0}   \cos(\phi/2)}{
P+2 \Delta ^2 t_{0}^2 \cos (\phi)
}\,,\\
F_{\rm 12}^{-}(\omega)&=\frac{2 i \omega \Delta  t_{0}   \sin(\phi/2)}{
P+2 \Delta ^2 t_{0}^2 \cos (\phi)}\,,
\end{split}
\end{equation}
where $\omega$  represents complex frequencies unless otherwise stated and  $P=\left(\Delta ^2-\omega ^2+\epsilon ^2\right)^2-2 t_{0}^2 \left(\omega ^2+\epsilon ^2\right)+t_{0}^4$.
\begin{figure}[!t]
\centering
	\includegraphics[width=0.95\columnwidth]{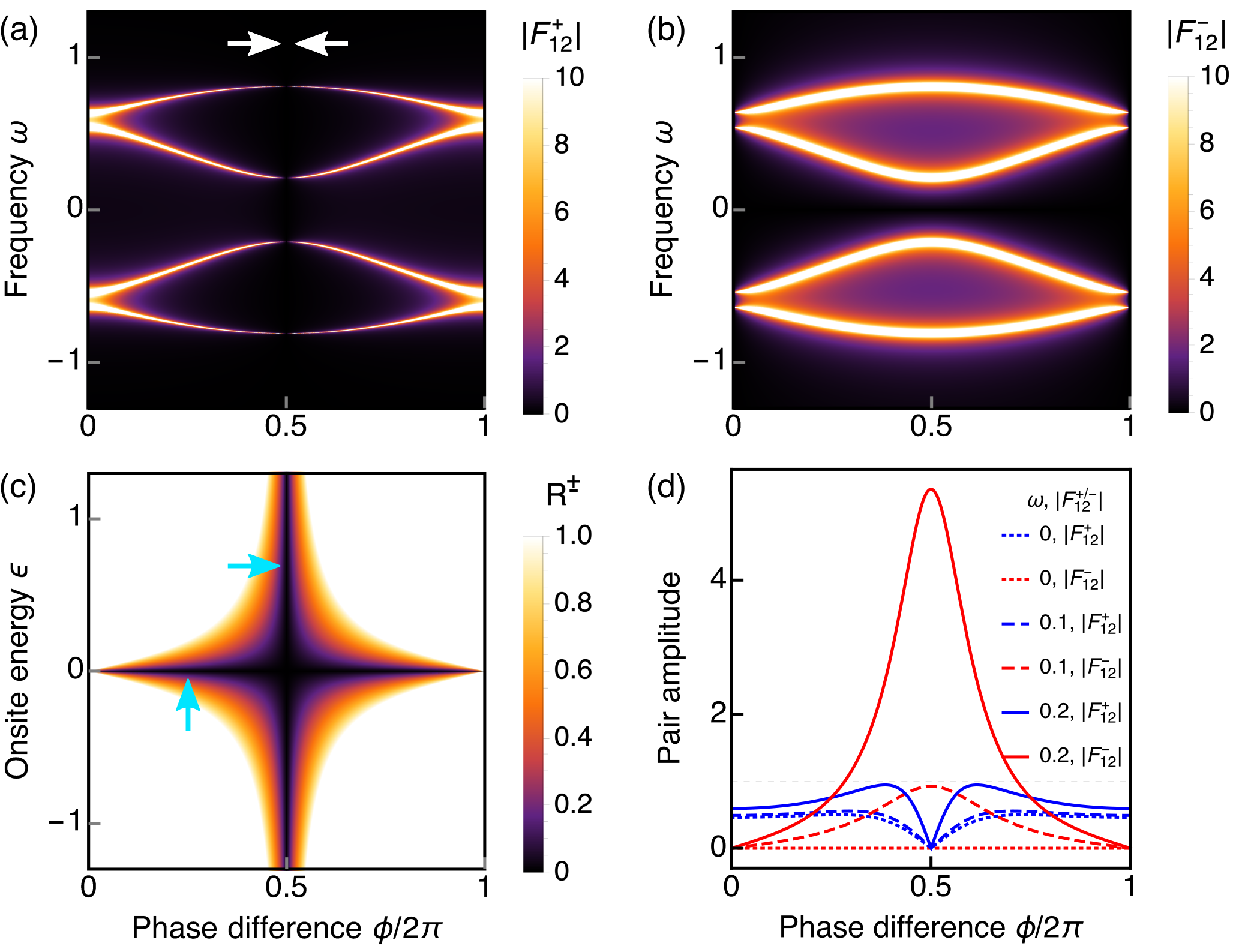}\\
 	\caption{(a,b) Symmetric even-$\omega$ and antisymmetric odd-$\omega$ nonlocal pair amplitudes ($F_{12}^{\pm}$)  in a JJ with two superconductors coupled directly as a function of $\omega$ and $\phi$ at $\epsilon=0.1$, with the color scale cut off at 10 for visualization. White arrows in (a) indicate that $|F_{12}^{+}|$ vanishes at $\phi=\pi$. (c) Ratio $R^{\pm}=|F_{12}^{+}|/|F_{12}^{-}|$ as a function of $\epsilon$ and $\phi$ at $\omega=0.2$. Cyan arrows indicate that $R$ vanishes either at $\epsilon=0$ or $\phi=\pi$. (d) Line cuts of (a,b) at fixed $\omega$. Parameters: $\Delta=0.5$, $t_{0}=0.3$.}
\label{Fig1} 
\end{figure}
First,  both pair amplitudes in Eqs.\,(\ref{LRLR})   have the same denominator which is an even function of $\omega$ and reveals the formation of Andreev bound states (ABSs) when $P+2 \Delta ^2 t_{0}^2 \cos (\phi)=0$. This is seen in the bright regions of Fig.\,\ref{Fig1},  where we plot the absolute value of the symmetric and antisymmetric amplitudes as a function of the phase difference $\phi$.  Second, the  numerators of both $F_{\rm 12}^{+}$   and $F_{\rm 12}^{-}$   have different functional dependences, oscillating with the phase difference $\phi$ in an alternate fashion as ${\rm cos}(\phi/2)$ and ${\rm sin}(\phi/2)$, respectively  \footnote{The dependence of the inter-superconductor pair amplitudes on the phase difference between two superconductors has been discussed before but only in the weak tunneling limit \cite{balatsky2018oddfrequency}, a regime that does not reveal the formation of ABSs in the pair amplitudes.}. While the  numerator of the symmetric term is an even function of $\omega$ with a linear dependence on  $\epsilon$, the antisymmetric component is interestingly linear in $\omega$ and, therefore, an odd function of frequency. The symmetric even-$\omega$ part vanishes either when $\epsilon=0$ or $\phi=\pi$, while the antisymmetric odd-$\omega$ pair amplitude remains remarkably  finite at these points  and even acquiring  large values. The surprising features of the nonlocal pair amplitudes can be seen by comparing the panels of Fig.\,\ref{Fig1}(a,b,d), where the vanishing values of the even-$\omega$ part is indicated by white arrows in Fig.\,\ref{Fig1}(a).   The vanishing values of the even-$\omega$ pairing can be better seen in  Fig.\,\ref{Fig1}(c) where we plot the ratio between the two pair amplitudes, $R^{\pm}=|F_{\rm 12}^{+}|/|F_{\rm 12}^{-}|=|(\epsilon/i\omega){\rm cot}(\phi/2)|$: 
$R^{\pm}$ vanishes either at $\epsilon=0$ or $\phi=\pi$. Note, however, that  since $F_{\rm 12}^{-}$ is an odd function of $\omega$ and thus vanishes at $\omega=0$, $R^{\pm}$ has a clear interpretation only for $\omega\neq0$. In sum, JJs with two superconductors exhibit highly tunable
 odd-$\omega$ pairing that is the only type of inter-superconductor pair correlations.  

\begin{figure}[!t]
\centering
	\includegraphics[width=0.95\columnwidth]{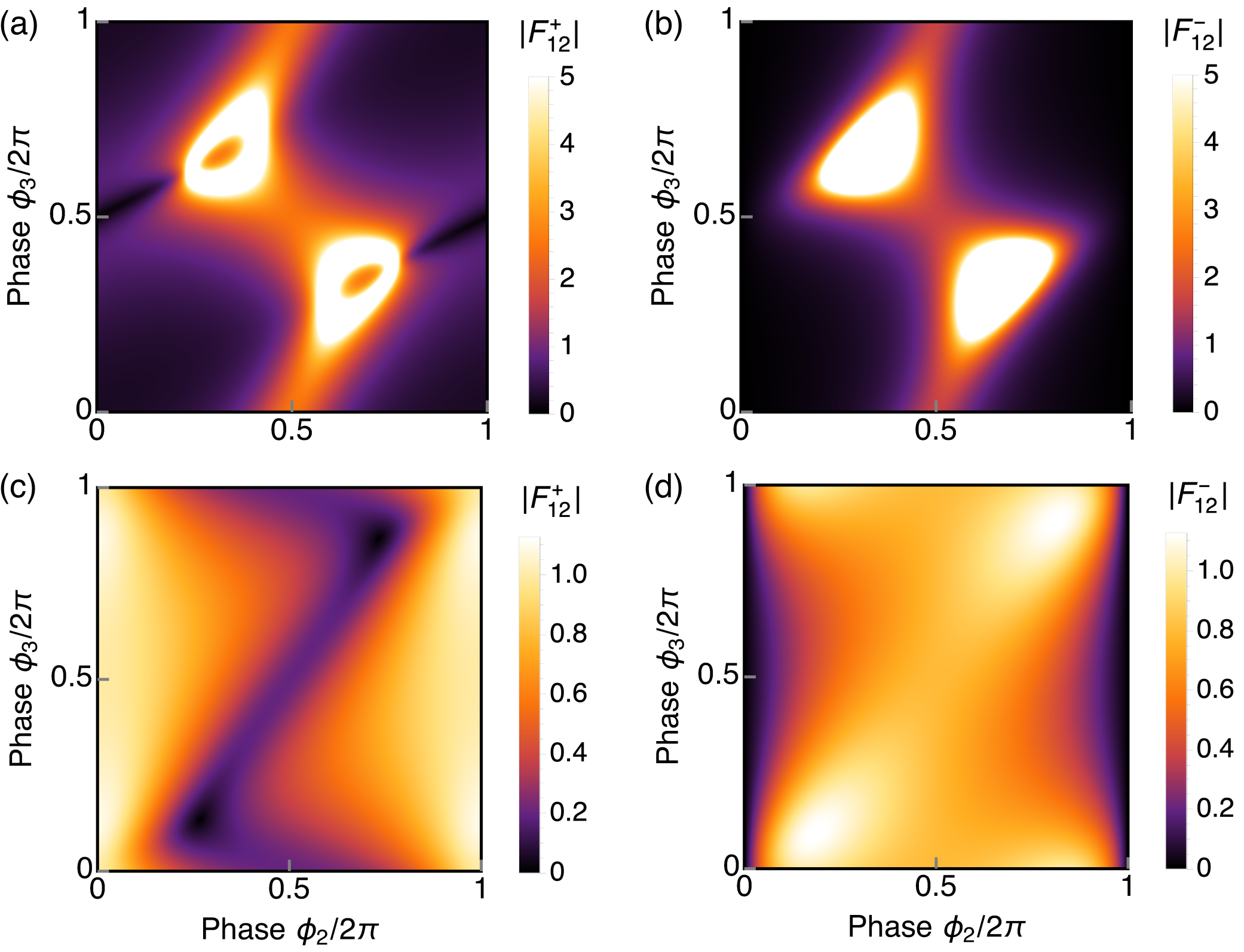}\\
 	\caption{(a,b) Symmetric even-$\omega$ ($F_{12}^{+}$) and antisymmetric  odd-$\omega$ ($F_{12}^{-}$) nonlocal pair amplitudes in a JJ with three superconductors coupled directly as a function of $\phi_{2}$ and $\phi_{3}$ at $\omega=0.1$ and $t_{0}=0.3$,   with the color scale cut off at 5 for visualization. (c,d) Same as in (a,b) but at  $\omega=1$ and $t_{0}=0.5$. Parameters: $\Delta=0.5$, $\epsilon=0$, $\phi_{1}=0$.}
\label{Fig2} 
\end{figure}

For JJs with more superconductors $n>2$, the expressions for the nonlocal pair amplitudes become lengthy, but still capturing the formation of ABSs in the denominator and with  numerators that strongly depend  on all $\phi_{i}$ \cite{SM}.  We find that the symmetric and antisymmetric pair amplitudes between nearest neighbour superconductors develop even- and odd-$\omega$ symmetries, respectively. While the odd-$\omega$ part is proportional to $\sim ({\rm e}^{i\phi_{j+1}}-{\rm e}^{i\phi_{j}})$,   the even-$\omega$ term  to $\sim ({\rm e}^{i\phi_{j+1}}+{\rm e}^{i\phi_{j}})+P(\phi_{1,\cdots,n})$, where $P$ is a function   of all the system parameters \cite{SM}.  Thus,   the odd-$\omega$ term depends on the ${\rm sine}$ of the phase difference of the involved superconductors as in JJs with two superconductors discussed above. However, the even-$\omega$ part has a ${\rm cosine}$ part as for JJs with two superconductors, but also an additional contribution due to the rest of the system. Nevertheless, both pair amplitudes exhibit a high degree of tunability by means of the superconducting phases. To visualize this fact, in Fig.\,\ref{Fig2} we plot the even-$\omega$ and odd-$\omega$ pair amplitudes for a JJ with three superconductors as a function of $\phi_{2}$ and $\phi_{3}$ at $\phi_{1}=0$. The main  feature of this figure is that the behaviour of both pair amplitudes is highly controllable by the superconducting phases. Interestingly, there are regions where the even-$\omega$ component acquires vanishing small values while the odd-$\omega$ remains sizeable large, see dark and bright  regions in Fig.\,\ref{Fig2}(a,c) and Fig.\,\ref{Fig2}(b,d), respectively.

The vanishing and finite values of the even- and odd-$\omega$ pair amplitudes can be further visualized in a simpler regime.  Specially,  for very weak couplings between superconductors $t_{0}$ and for superconductors with the same onsite energy $\epsilon$, the nearest neighbour nonlocal pair amplitudes up to linear order in $t_{0}$ are given by \cite{SM}
\begin{equation}
\label{FJJ}
\begin{split}
F_{j,j+1}^{+}(\omega)&\approx\frac{ \epsilon \Delta  t_{0}  \left(e^{i \phi_{j+1}}+e^{i \phi_{j}}\right) }{
(\Delta^{2}-\omega^{2}+\epsilon^{2})^{2}
}\,,\\
F_{j,j+1}^{-}(\omega)&\approx\frac{\omega \Delta  t_{0} \left(e^{i \phi_{j+1}}-e^{i \phi_{j}}\right)}{
(\Delta^{2}-\omega^{2}+\epsilon^{2})^{2}}\,,
\end{split}
\end{equation}
where $j=1,\cdots, n$ and $\phi_{n+1}=\phi_{1}$. Strikingly, only the  pair amplitudes  between nearest neighbour superconductors remain finite at leading order in $t_{0}$  \footnote{We have verified  that higher order in $t_{0}$ produces finite values of both even- and odd-$\omega$ pair amplitudes between next nearest neighbour superconductors.}. 
As expected, $F_{j,j+1}^{+}$ and $F_{j,j+1}^{-}$ in Eqs.\,(\ref{FJJ}) exhibit even- and odd-$\omega$  spin-singlet symmetries, respectively. Interestingly,   both pair amplitudes acquire the same form as their counterparts in JJs with two superconductors, see Eqs.\,(\ref{LRLR}). In this regime, the even-$\omega$ pairing thus vanishes either at $\epsilon=0$ or when  $e^{i \phi_{j+1}}+e^{i \phi_{j}}=0$ which needs a phase difference of $\phi_{j+1}-\phi_{j}=\pi$ between superconductors. However, the odd-$\omega$ component remains always finite in this regime, exhibiting  high tunability by $\phi_{j}$.  We have verified that this behaviour remains even in JJs with finite superconductors and also in JJs with superconductors coupled via a normal region \cite{SM}.   Hence, multi-superconductor JJs  represent a rich platform for the generation and control of nonlocal odd-$\omega$ pair correlations that do not require magnetic elements. Before closing this part, we highlight that the odd-$\omega$ pair amplitudes presented here are a proximity-induced superconducting effect bound to the device, exhibiting  wide controllability by the  superconducting phases and with important impact on physical observables as we discuss next.


\section{CAR detection of  odd-$\omega$ pairing}
\label{sectionV}
Having established the emergence of  inter-superconductor odd-$\omega$  pairs in multi-superconductor JJs, now we inspect a direct detection protocol. Due to the nonlocal character of the pair correlations found here, it is natural to explore nonlocal transport of Cooper pairs \cite{PhysRevB.93.201402,PhysRevB.97.075408,PhysRevB.92.100507,PhysRevB.106.L100502}. Without loss of generality, we   focus on JJs  formed by two superconductors and aim at detecting the odd-$\omega$ pairs obtained in Eqs.\,\ref{LRLR}. Hence, we  attach two normal leads at the left and the right of the system as in Fig.\,\ref{Fig0} and include them in our model via  retarded self-energies $\Sigma_{\rm L(R)}^{r}$, such that the system's retarded Green's function is  $G^{r}(\omega)=(\omega+i0^{+}-H_{\rm 2JJ}-\Sigma^{r}_{\rm L}-\Sigma_{\rm R}^{r})^{-1}$ \cite{datta1997electronic}. Here, $H_{\rm 2JJ}$  describes the JJ described by Eq.\,(\ref{JJEqs}) with $n=2$   and $\omega$ now  represents real frequencies.  In the wide-band limit,  $\Sigma^{r}_{j}=-i\Gamma_{j}/2$, where $\Gamma_{j}=\pi |\tau|^{2}\rho_{j}$ characterizes the coupling to  lead  $j$   with  surface density of states $\rho_{j}$, and $\tau$ the hopping between leads and superconductors. 

At weak $\Gamma_{j}$, the   JJ  can be probed by nonlocal transport. Specially, the transport of Cooper pairs is characterized by  nonlocal Andreev reflection or crossed  Andreev reflection ($T_{\rm CAR}$), which competes with electron tunneling ($T_{\rm ET}$) to  determine the nonlocal   conductance $\sim  (T_{\rm CAR}-T_{\rm ET})$ \cite{SM}. These CAR and ET processes involve electron-hole (hole-electron) and electron-electron (hole-hole) transfers, $T_{\rm CAR}=T_{eh}+T_{he}$  and $T_{\rm ET}=T_{ee}+T_{hh}$, which can be obtained from $G^{r}$ as \cite{PhysRevB.93.201402}
\begin{equation}
\begin{split}
T_{ee}&=\Gamma_{\rm L}^{e}\Gamma_{\rm R}^{e}|{g}^{r}_{12}|^{2}\,,
T_{hh}=\Gamma_{\rm L}^{h}\Gamma_{\rm R}^{h}|\bar{g}^{r}_{12}|^{2}\,,\\
T_{eh}&=\Gamma_{\rm L}^{e}\Gamma_{\rm R}^{h}|F_{12}^{r}|^{2}\,,
T_{he}=\Gamma_{\rm L}^{h}\Gamma_{\rm R}^{e}|\bar{F}_{12}^{r}|^{2}\,,
\end{split}
\end{equation}
where $g^{r}_{12}$ ($\bar{g}^{r}_{12}$) and $F_{12}^{r}$ ($\bar{F}_{12}^{r}$) are the normal and anomalous (or pair amplitude) components of the inter-superconductor retarded Green's function, obtained from $G^{r}$ \cite{SM}. Interestingly, the CAR processes $T_{eh(he)}$ are directly determined by the squared   modulus of the inter-superconductor pair amplitudes $F_{12}^{r}$.  We note that, while the pair amplitudes $F_{12}^{r}$ and $\bar{F}_{12}^{r}$ are not directly measurable, their modulo respectively determines the finite value of the nonlocal probabilities $T_{eh}$ and $T_{he}$, thus facilitating the detection of these emergent pairings.

Under general circumstances,  $F_{12}^{r}$ includes both symmetric even-$\omega$ and antisymmetric odd-$\omega$ terms, the symmetric part vanishes at $\epsilon=0$ for any $\phi$, see Eqs.\,\ref{LRLR}. Thus, the CAR amplitudes have the potential to directly probe the antisymmetric inter-superconductor odd-$\omega$ pairing. However, as shown above, the CAR processes $T_{eh (he)}$ are always accompanied by electron tunnelings $T_{ee(hh)}$. Therefore, even if $T_{eh (he)}$ directly probes odd-$\omega$ pairs, their total effect in the non-local conductance  can be masked if $T_{ee(hh)}$ are larger. For this reason, to directly detect inter-superconductor odd-$\omega$ pairing, a regime where $T_{ee(hh)}\ll T_{eh(he)}$ is needed. Even though  this regime might sound challenging to find,  we now demonstrate that it is in fact possible. To show this, we  consider $\phi_{1}=-\phi/2$, $\phi_{2}=\phi/2$ and   assume symmetric couplings to the leads $\Gamma_{j}=\Gamma$. Then, for $\epsilon=0$, $g^{r}_{12}$ and the antisymmetric pair amplitude $F_{12}^{r,-}$  are given by \cite{SM}
\begin{equation}
\begin{split}
g_{\rm 12}^{r}&=-4t_{0}\{(\Gamma-2i\omega)^{2}+4t_{0}^{2}+4\Delta^{2}{\rm e}^{-i\phi}\}/D\,,\\
F_{\rm 12}^{r,-}&=16it_{0}\Delta(2\omega+i\Gamma){\rm sin}(\phi/2)/D\,,
\end{split}
\end{equation}
where $D=16t_{0}^{4}+[4\Delta^{2}+(\Gamma-2i\omega)^{2}]^{2}+8t_{0}^{2}(\Gamma-2i\omega)^{2}+32t_{0}^{2}\Delta^{2}{\rm cos}(\phi)$, and $\bar{g}^{r}_{12}(\phi)=-g_{\rm 12}^{r}(-\phi)$, and $\bar{F}^{r}_{12}(\phi)=F_{12}^{r}(\phi)$. 
We note that $F_{12}^{r}$ can be obtained from Eqs.\,(\ref{LRLR}) by replacing $\omega\rightarrow\omega+i0^{+}+i\Gamma/2$. 

\begin{figure}[!t]
\centering
	\includegraphics[width=0.95\columnwidth]{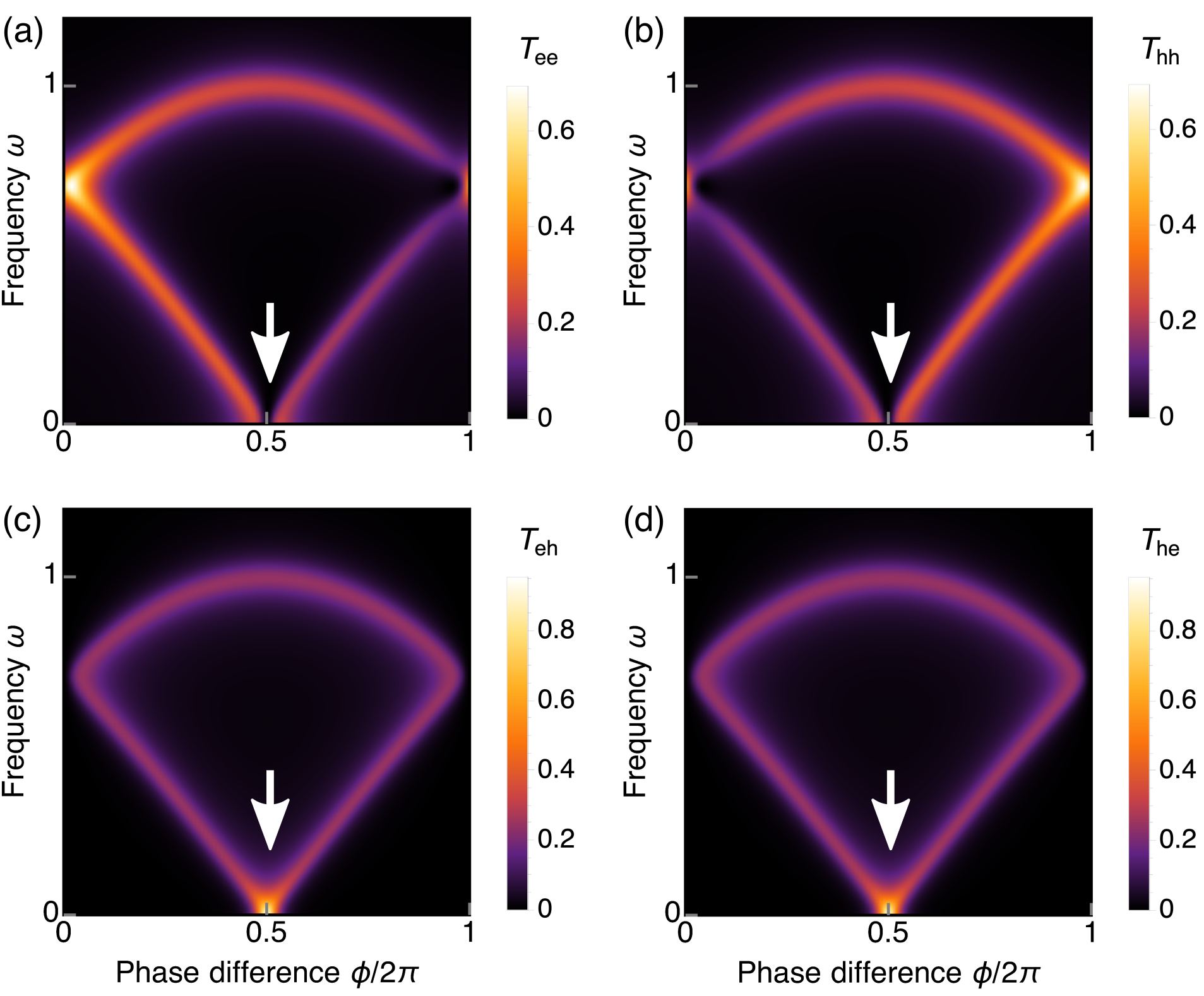}\\
 	\caption{Electron tunneling (top row) and crossed Andreev reflection (bottom) processes as a function of $\omega$ and $\phi$. Parameters: $\Delta=0.5$, $\epsilon=0$, $\Gamma_{j}=0.1$, $t_{0}=0.5$.}
\label{Fig3} 
\end{figure}
Now, we can exploit the fact that the energy of the ABSs  at $\epsilon=0$ and $\phi=\pi$ is given by $|\omega_{\pm}|=|t_{0}-\Delta|$, which clearly vanishes for $t_{0}=\Delta$. In this regime we have $|g_{\rm 12}^{r}|/|F_{\rm 12}^{r,-}|\approx\omega/(2\Delta)\ll1$    for low frequencies. Thus, it is possible to obtain a regime where the antisymmetric pair amplitude is larger than the normal contribution. Hence, in this regime $T_{eh(he)}$ are expected to be larger than $T_{ee(hh)}$ and constitute the main contribution to the non-local conductance, whose finite value indicates a direct evidence of inter-superconductor odd-$\omega$ pairing. 

To visualize the above argument, in Fig.\,\ref{Fig3} we plot ET and CAR processes as a function of $\phi$ and $\omega$ at $\epsilon=0$. The most important feature is that at high frequencies, ET processes $T_{ee(hh)}$ acquire large values near $\phi=0,2\pi$ but are vanishing small at low $\omega$ near $\phi=\pi$, in line with the discussion presented above. Interestingly, the CAR processes $T_{eh(he)}$ acquire large values around $\phi=\pi$ at low frequencies but smaller values at higher frequencies. The finite values of these CAR processes directly probe the formation of induced odd-$\omega$ pairs. Of particular relevance here are the values around $\phi=\pi$ and low $\omega$, because, at such points, CAR dominates over ET and   it thus determines the nonlocal conductance.  We have verified that  this behaviour also holds for JJs with more than two superconductors but in the weak tunneling regime, thus supporting the direct detection  of proximity-induced inter-superconductor odd-$\omega$ pairing in a nonlocal transport measurement. Hence, despite being an induced effect, the nonlocal odd-$\omega$ pairs determine CAR processes by simply tuning the superconducting phases in multi-superconductor JJs.

\section{Conclusions}
\label{sectionVI}
In conclusion, we have studied multi-superconductor Josephson junctions and found that inter-superconductor even- and odd-$\omega$ Cooper pairs can be generated, controlled, and detected by virtue of the superconducting phases. We found that even-$\omega$ pairing vanishes when the   phase differences between two superconductors is $\pi$, thus leaving only odd-$\omega$ pairing as the only type of inter-superconductor  pair correlations. While this finding is exact for Josephson junctions with two superconductors, it is only valid at   weak couplings between superconductors in junctions with more than two superconductors. Due to the vanishing of even-$\omega$ pairing, only odd-$\omega$ pairs contribute to CAR processes, whose finite values directly probe the presence of odd-$\omega$ Cooper pairs.

Given the advances in the fabrication of superconducting heterostructures, including a promising tunability of CAR processes~\cite{PhysRevX.13.031031}, we expect that the physics discussed here could be soon realized in multi-terminal Josephson junctions \cite{PhysRevX.10.031051,strambini2016omega,draelos2019supercurrent,graziano2022selective,coraiola2023hybridisation} and in superconducting quantum dots  \cite{van2006supercurrent,hofstetter2009cooper,de2010hybrid,dirks2011transport,lee2014spin,pillet2019nonlocal,zitko2019superconducting,bargerbos2022spectroscopy,dvir2023realization,bordin2023crossed}. Of particular relevance are   Refs.\,\cite{PhysRevX.10.031051,strambini2016omega,draelos2019supercurrent,graziano2022selective,coraiola2023hybridisation} because they have already demonstrated the fabrication of multi-superconductor Josephson junctions and the control of several superconducting phases.  
 In this regard, our work offers an entirely unexplored route for the generation, control, and detection of   odd-$\omega$ Cooper pairs that might be even possible to explore using already existent experimental techniques.


\section{Acknowledgements}
We thank  Y. Asano, S. Ikegaya,  and S. Tamura  for insightful discussions.   
J. C. acknowledges financial support from the Japan Society for the Promotion of Science via the International Research Fellow Program, from the Swedish Research Council  (Vetenskapsr\aa det Grant No.~2021-04121), and from Royal Swedish Academy of Sciences (Grant No.~PH2022-0003).
P. B. acknowledges support from the Spanish CM ``Talento Program'' project No.~2019-T1/IND-14088 and No.~2023-5A/IND-28927, and the Agencia Estatal de Investigaci\'on project No.~PID2020-117992GA-I00 and No.~CNS2022-135950. Y. T. acknowledges support from JSPS with Grants-in-Aid for Scientific Research (A) (KAKENHI Grant No. JP20H00131), Grants-in-Aid for Scientific Research (B) (KAKENHI Grant No. JP20H01857), Grants-in-Aid for Scientific Research (C) (KAKENHI Grants  No. 23K17668 and 24K00583), and the JSPS Core-to-Core program ``Oxide Superspin'' international network.

\bibliography{biblio}
\onecolumngrid



\section*{Supplementary material}

# Supplemental Material for "Controllable odd-frequency Cooper pairs in multi-superconductor Josephson junctions"

Jorge Cayao,[1] Pablo Burset,[2] and Yukio Tanaka[3]

[1]*Department of Physics and Astronomy, Uppsala University, Box 516, S-751 20 Uppsala, Sweden*
[2]*Department of Theoretical Condensed Matter Physics,*
*Condensed Matter Physics Center (IFIMAC) and Instituto Nicolás Cabrera,*
*Universidad Autónoma de Madrid, 28049 Madrid, Spain*
[3]*Department of Applied Physics, Nagoya University, Nagoya 464–8603, Japan*
(Dated: April 12, 2024)

In this supplemental material we present a symmetry classification of pair correlations in multi-superconductor Josephson junctions. We also provide all the necessary steps for the calculation of the Green's function of a Josephson junction with multiple superconductors coupled directly or through a normal region, and the nonlocal transport properties of such junctions. Finally, we also obtain the nonlocal pair correlations in finite length Josephson junctions.

## ALLOWED PAIR SYMMETRIES IN MULTI-SUPERCONDUCTOR JOSEPHSON JUNCTIONS

Before carrying out any calculation in a specific system, here we present all the allowed superconducting pair symmetries in multi-superconductor Josephson junctions with quantum numbers that involve frequency ($\omega$), spins ($\sigma, \sigma'$), superconductor indices ($n, m$), and spatial coordinates ($x, x'$). For this purpose, we remind that the antisymmetry condition dictates that the pair amplitudes $F_{n,m}^{\sigma,\sigma'}(\omega; x, x')$ must be antisymmetric under the total exchange of quantum numbers, namely,

$$F_{n,m}^{\sigma,\sigma'}(\omega; x, x') = -F_{m,n}^{\sigma',\sigma}(-\omega; x', x), \tag{S1}$$

where $\omega$ represents complex frequencies unless otherwise stated. Thus, the allowed pair symmetries should fulfil Eq. (S1) under the total exchange of quantum numbers. However, $F_{n,m}^{\sigma,\sigma'}(\omega; x, x')$ can be even or odd under the individual exchange of quantum numbers as long as Eq. (S1) holds. Thus, for instance, the pair amplitude can be even (odd) under the exchange of frequency $\omega \to -\omega$ and pick up a plus (minus) sign, which translates as

$$F_{n,m}^{\sigma,\sigma'}(\omega; x, x') = \pm F_{n,m}^{\sigma,\sigma'}(-\omega; x, x'). \tag{S2}$$

Moreover, as stated at the beginning of this part, $F_{n,m}^{\sigma,\sigma'}(\omega; x, x')$ can be even (odd) under the individual exchange of the rest of the quantum numbers. Thus, $F_{n,m}^{\sigma,\sigma'}(\omega; x, x')$ can be even (odd) under the exchange of spins, superconducting indices, or spatial coordinates, respectively, when

$$\begin{aligned} F_{n,m}^{\sigma,\sigma'}(\omega; x, x') &= \pm F_{n,m}^{\sigma',\sigma}(\omega; x, x'), \\ F_{n,m}^{\sigma,\sigma'}(\omega; x, x') &= \pm F_{m,n}^{\sigma,\sigma'}(\omega; x, x'), \\ F_{n,m}^{\sigma,\sigma'}(\omega; x, x') &= \pm F_{n,m}^{\sigma,\sigma'}(\omega; x', x). \end{aligned} \tag{S3}$$

Therefore, the allowed pair amplitudes can be obtained by performing all the possible combinations of the individual exchanges of quantum numbers (Eqs. (S2) and (S3)) that fulfil the antisymmetry condition Eq. (S1). By doing this, we find eight allowed pair symmetry classes that respect the antisymmetry condition, where 4 correspond to odd-frequency correlations, see Table S1. Of particular relevance is that the superconducting index (sup. index) plays a crucial role for broadening the allowed pair symmetries. Table S1 is presented as Table 1 in the section on "*Inter-superconductor pair amplitudes in JJs*" of the main text.

In the absence of any spin-mixing field, the spin symmetry of the emergent pair correlations is the same as the spin symmetry of the parent superconductor. Thus, the pair symmetry classes allowed in our study, where no spin-mixing field is present, are ESEE and OSOE pair symmetry classes: they correspond to a pair amplitude with even-frequency (odd-frequency) spin-singlet even (odd) in superconductor indices, even parity. By including a spin-mixing field, it is possible to obtain odd-frequency spin-triplet pair amplitudes which correspond to OTEE and OTOO pair symmetry classes in Table S1, which could be used as sources of spins highly controllable by the superconducting phases and thus promising for superconducting spintronics. Since we do not have spin-mixing fields in the results presented in the main text, the pair symmetries therein exhibit the spin-symmetry of the parent superconductor, namely, spin singlet. This is specially discussed in the section on "*Inter-superconductor pair amplitudes in JJs*" of the main text.

| Pair symmetries in multi-superconductor Josephson junctions | | | | |
|---|---|---|---|---|
| Frequency $(\omega \leftrightarrow -\omega)$ | Spin $(\uparrow \leftrightarrow \downarrow)$ | Sup. index $(n \leftrightarrow m)$ | Parity $(x \leftrightarrow x')$ | Pair symmetry class (total exchange) |
| **E**ven | **S**inglet | **E**ven | **E**ven | ESEE |
| **E**ven | **S**inglet | **O**dd | **O**dd | ESOO |
| **E**ven | **T**riplet | **E**ven | **O**dd | ETEO |
| **E**ven | **T**riplet | **O**dd | **E**ven | ETOE |
| **O**dd | **S**inglet | **E**ven | **O**dd | OSEO |
| **O**dd | **S**inglet | **O**dd | **E**ven | OSOE |
| **O**dd | **T**riplet | **E**ven | **E**ven | OTEE |
| **O**dd | **T**riplet | **O**dd | **O**dd | OTOO |

TABLE S1. Allowed superconducting pair symmetries in multi-superconductor Josephson junctions under the presence of spin-mixing fields. The classes ESEE and OSOE, which are spin-singlet, correspond to the pair correlations reported in the main text of this work.

## GREEN'S FUNCTIONS OF JOSEPHSON JUNCTIONS WITH SUPERCONDUCTORS COUPLED DIRECTLY

To obtain the Green's functions of the Josephson junctions studied in the main text, we first write their model Hamiltonian in Nambu (electron-hole) space and then obtain them from the equation of motion $[\omega - H_{\rm nJJ}]G(\omega) = I$, where $H_{\rm nJJ}$ is the Hamiltonian of the phase-biased Josephson junctions and $G(\omega)$ its associated Green's function. A Josephson junction with $n$ superconductors coupled directly is modeled by Eq. (1) in the main text. In Nambu space, the Hamiltonian of each superconductor S$_j$ is given by

$$H_j = \begin{pmatrix} \epsilon_j & \Delta e^{i\phi_j} \\ \Delta e^{-i\phi_j} & -\epsilon_j \end{pmatrix}, \tag{S4}$$

where $\epsilon_j$ represents the onsite energy of the superconductor, $\Delta$ is the spin singlet $s$-wave pair potential and $\phi_j$ its superconducting phase. Similarly, the coupling between superconductors is described by

$$V_{j,j+1} = \begin{pmatrix} t_{j,j+1} & 0 \\ 0 & -t_{j,j+1} \end{pmatrix}, \tag{S5}$$

where $t_{j,j+1}$ represents the coupling strength between nearest superconductors S$_j$ and S$_{j+1}$. For simplicity, we consider that all such couplings are the same $t_{j,j+1} = t_0$ and thus drop the indices in the coupling matrix ($V_{j,j+1} = V$). Below, we discuss Josephson junctions with distinct superconductors and obtain expressions for their associated Green's functions.

### Josephson junctions with two superconductors

The Hamiltonian of a Josephson junction between two superconductors is

$$H_{\rm 2JJ}^{(1)} = \begin{pmatrix} H_1 & V \\ V^\dagger & H_2 \end{pmatrix}. \tag{S6}$$

The eigenvalues of this Hamiltonian are given by

$$E_j = \pm \frac{1}{\sqrt{2}} \sqrt{2\Delta^2 + 2t_0^2 + \epsilon_1^2 + \epsilon_2^2 \pm \sqrt{4t_0^2 \left(2\Delta^2 + (\epsilon_1 + \epsilon_2)^2\right) - 8\Delta^2 t_0^2 \cos(\phi) + (\epsilon_1^2 - \epsilon_2^2)^2}}, \tag{S7}$$

which, for $\epsilon_{1,2} = \epsilon$, reduce to

$$E_j = \pm \sqrt{\Delta^2 + t_0^2 + \epsilon^2 \pm \sqrt{2t_0^2 \left[2\epsilon^2 + \Delta^2(1 - \cos(\phi))\right]}}. \tag{S8}$$

Then,

$$\begin{aligned} E(\phi = 0) &= \pm \sqrt{(t_0 \pm |\epsilon|)^2 + \Delta^2}, \\ E(\phi = \pi) &= \pm |t_0 \pm \sqrt{\Delta^2 + \epsilon^2}|. \end{aligned} \tag{S9}$$

Therefore, the gap closes at $\phi = \pi$ if $t_0 = \sqrt{\Delta^2 + \epsilon^2}$; for $\epsilon = 0$, the gap closes when $t_0 = \Delta$. The associated Green's function has the following structure

$$G_{2JJ} = [\omega - H_{2JJ}]^{-1} = \begin{pmatrix} G_{11} & G_{12} \\ G_{21} & G_{22} \end{pmatrix}. \tag{S10}$$

The entries of $G_{2JJ}$ correspond to the Green's functions inside the superconductors ($G_{11(22)}$) or between the superconductors ($G_{12(21)}$). We thus term these components as intra- and inter-superconductor Green's functions, respectively. The Nambu structure of each $G_{ij}$ is given by

$$G_{ij} = \begin{pmatrix} G_{0,ij} & F_{ij} \\ \bar{F}_{ij} & \bar{G}_{0,ij} \end{pmatrix}, \tag{S11}$$

where the diagonal elements allow us to obtain the density of states, while the off-diagonal ones give the pair amplitudes. Thus, for the pair amplitudes we obtain

$$
\begin{aligned}
F_{11} &= -\frac{\Delta e^{i(\phi_1+\phi_2)}\left[e^{i\phi_1}\left(\Delta^2 - \omega^2 + \epsilon_2^2\right) + t_0^2 e^{i\phi_2}\right]}{e^{i(\phi_1+\phi_2)}\left[\left(\Delta^2 - \omega^2 + \epsilon_1^2\right)\left(\Delta^2 - \omega^2 + \epsilon_2^2\right) - 2t_0^2\left(\omega^2 + \epsilon_1\epsilon_2\right) + t_0^4\right] + \Delta^2 t_0^2\left(e^{2i\phi_1} + e^{2i\phi_2}\right)}, \\
\bar{F}_{11} &= -\frac{\Delta\left(e^{i\phi_2}\left(\Delta^2 - \omega^2 + \epsilon_2^2\right) + t_0^2 e^{i\phi_1}\right)}{e^{i(\phi_1+\phi_2)}\left(\left(\Delta^2 - \omega^2 + \epsilon_1^2\right)\left(\Delta^2 - \omega^2 + \epsilon_2^2\right) - 2t_0^2\left(\omega^2 + \epsilon_1\epsilon_2\right) + t_0^4\right) + \Delta^2 t_0^2 e^{2i\phi_1} + \Delta^2 t_0^2 e^{2i\phi_2}}, \\
F_{22} &= -\frac{\Delta e^{i(\phi_1+\phi_2)}\left[e^{i\phi_2}\left(\Delta^2 - \omega^2 + \epsilon_1^2\right) + t_0^2 e^{i\phi_1}\right]}{e^{i(\phi_1+\phi_2)}\left[\left(\Delta^2 - \omega^2 + \epsilon_1^2\right)\left(\Delta^2 - \omega^2 + \epsilon_2^2\right) - 2t_0^2\left(\omega^2 + \epsilon_1\epsilon_2\right) + t_0^4\right] + \Delta^2 t_0^2\left(e^{2i\phi_1} + e^{2i\phi_2}\right)}, \\
\bar{F}_{22} &= -\frac{\Delta\left(e^{i\phi_1}\left(\Delta^2 - \omega^2 + \epsilon_1^2\right) + t_0^2 e^{i\phi_2}\right)}{e^{i(\phi_1+\phi_2)}\left(\left(\Delta^2 - \omega^2 + \epsilon_1^2\right)\left(\Delta^2 - \omega^2 + \epsilon_2^2\right) - 2t_0^2\left(\omega^2 + \epsilon_1\epsilon_2\right) + t_0^4\right) + \Delta^2 t_0^2 e^{2i\phi_1} + \Delta^2 t_0^2 e^{2i\phi_2}}, \\
F_{12} &= \frac{\Delta\, t_0\, e^{i(\phi_1+\phi_2)}\left(\epsilon_1 e^{i\phi_2} + \epsilon_2 e^{i\phi_1}\right) + \Delta\, t_0\, \omega\, e^{i(\phi_1+\phi_2)}\left(e^{i\phi_2} - e^{i\phi_1}\right)}{e^{i(\phi_1+\phi_2)}\left[\left(\Delta^2 - \omega^2 + \epsilon_1^2\right)\left(\Delta^2 - \omega^2 + \epsilon_2^2\right) - 2t_0^2\left(\omega^2 + \epsilon_1\epsilon_2\right) + t_0^4\right] + \Delta^2 t_0^2\left(e^{2i\phi_1} + e^{2i\phi_2}\right)}, \\
\bar{F}_{12} &= \frac{\Delta t_0\left(e^{i\phi_1}(\epsilon_1 - \omega) + e^{i\phi_2}(\omega + \epsilon_2)\right)}{e^{i(\phi_1+\phi_2)}\left(\left(\Delta^2 - \omega^2 + \epsilon_1^2\right)\left(\Delta^2 - \omega^2 + \epsilon_2^2\right) - 2t_0^2\left(\omega^2 + \epsilon_1\epsilon_2\right) + t_0^4\right) + \Delta^2 t_0^2 e^{2i\phi_1} + \Delta^2 t_0^2 e^{2i\phi_2}}, \\
F_{21} &= \frac{\Delta\, t_0\, e^{i(\phi_1+\phi_2)}\left(\epsilon_1 e^{i\phi_2} + \epsilon_2 e^{i\phi_1}\right) + \Delta\, t_0\, \omega\, e^{i(\phi_1+\phi_2)}\left(e^{i\phi_1} - e^{i\phi_2}\right)}{e^{i(\phi_1+\phi_2)}\left[\left(\Delta^2 - \omega^2 + \epsilon_1^2\right)\left(\Delta^2 - \omega^2 + \epsilon_2^2\right) - 2t_0^2\left(\omega^2 + \epsilon_1\epsilon_2\right) + t_0^4\right] + \Delta^2 t_0^2\left(e^{2i\phi_1} + e^{2i\phi_2}\right)}, \\
\bar{F}_{21} &= \frac{\Delta t_0\left(e^{i\phi_1}(\omega + \epsilon_1) + e^{i\phi_2}(\epsilon_2 - \omega)\right)}{e^{i(\phi_1+\phi_2)}\left(\left(\Delta^2 - \omega^2 + \epsilon_1^2\right)\left(\Delta^2 - \omega^2 + \epsilon_2^2\right) - 2t_0^2\left(\omega^2 + \epsilon_1\epsilon_2\right) + t_0^4\right) + \Delta^2 t_0^2 e^{2i\phi_1} + \Delta^2 t_0^2 e^{2i\phi_2}}.
\end{aligned}
\tag{S12}
$$

For the normal components we get

$$
\begin{aligned}
G_{0,11} &= -\frac{e^{i(\phi_1+\phi_2)}\left((\omega + \epsilon_1)\left(\Delta^2 - \omega^2 + \epsilon_2^2\right) + t_0^2(\omega - \epsilon_2)\right)}{e^{i(\phi_1+\phi_2)}\left(\left(\Delta^2 - \omega^2 + \epsilon_1^2\right)\left(\Delta^2 - \omega^2 + \epsilon_2^2\right) - 2t_0^2\left(\omega^2 + \epsilon_1\epsilon_2\right) + t_0^4\right) + \Delta^2 t_0^2 e^{2i\phi_1} + \Delta^2 t_0^2 e^{2i\phi_2}}, \\
\bar{G}_{0,11} &= \frac{e^{i(\phi_1+\phi_2)}\left((\epsilon_1 - \omega)\left(\Delta^2 - \omega^2 + \epsilon_2^2\right) - t_0^2(\omega + \epsilon_2)\right)}{e^{i(\phi_1+\phi_2)}\left(\left(\Delta^2 - \omega^2 + \epsilon_1^2\right)\left(\Delta^2 - \omega^2 + \epsilon_2^2\right) - 2t_0^2\left(\omega^2 + \epsilon_1\epsilon_2\right) + t_0^4\right) + \Delta^2 t_0^2 e^{2i\phi_1} + \Delta^2 t_0^2 e^{2i\phi_2}}, \\
G_{0,22} &= \frac{e^{i(\phi_1+\phi_2)}\left(t_0^2(\epsilon_1 - \omega) - (\omega + \epsilon_2)\left(\Delta^2 - \omega^2 + \epsilon_1^2\right)\right)}{e^{i(\phi_1+\phi_2)}\left(\left(\Delta^2 - \omega^2 + \epsilon_1^2\right)\left(\Delta^2 - \omega^2 + \epsilon_2^2\right) - 2t_0^2\left(\omega^2 + \epsilon_1\epsilon_2\right) + t_0^4\right) + \Delta^2 t_0^2 e^{2i\phi_1} + \Delta^2 t_0^2 e^{2i\phi_2}}, \\
\bar{G}_{0,22} &= \frac{e^{i(\phi_1+\phi_2)}\left((\epsilon_2 - \omega)\left(\Delta^2 - \omega^2 + \epsilon_1^2\right) - t_0^2(\omega + \epsilon_1)\right)}{e^{i(\phi_1+\phi_2)}\left(\left(\Delta^2 - \omega^2 + \epsilon_1^2\right)\left(\Delta^2 - \omega^2 + \epsilon_2^2\right) - 2t_0^2\left(\omega^2 + \epsilon_1\epsilon_2\right) + t_0^4\right) + \Delta^2 t_0^2 e^{2i\phi_1} + \Delta^2 t_0^2 e^{2i\phi_2}}, \\
G_{0,12} &= \frac{t_0\left(-\Delta^2 e^{i(\phi_1-\phi_2)} - t_0^2 + (\omega + \epsilon_1)(\omega + \epsilon_2)\right)}{\left(\Delta^2 - \omega^2 + \epsilon_1^2\right)\left(\Delta^2 - \omega^2 + \epsilon_2^2\right) - 2t_0^2\left(\Delta^2(-\cos(\phi_1 - \phi_2)) + \omega^2 + \epsilon_1\epsilon_2\right) + t_0^4}, \\
\bar{G}_{0,12} &= \frac{t_0 e^{i(\phi_1+\phi_2)}\left(t_0^2 + (\epsilon_1 - \omega)(\omega - \epsilon_2)\right) + \Delta^2 t_0 e^{2i\phi_2}}{e^{i(\phi_1+\phi_2)}\left(\left(\Delta^2 - \omega^2 + \epsilon_1^2\right)\left(\Delta^2 - \omega^2 + \epsilon_2^2\right) - 2t_0^2\left(\omega^2 + \epsilon_1\epsilon_2\right) + t_0^4\right) + \Delta^2 t_0^2 e^{2i\phi_1} + \Delta^2 t_0^2 e^{2i\phi_2}}, \\
G_{0,21} &= \frac{t_0\left(-\Delta^2 e^{-i(\phi_1-\phi_2)} - t_0^2 + (\omega + \epsilon_1)(\omega + \epsilon_2)\right)}{\left(\Delta^2 - \omega^2 + \epsilon_1^2\right)\left(\Delta^2 - \omega^2 + \epsilon_2^2\right) - 2t_0^2\left(\Delta^2(-\cos(\phi_1 - \phi_2)) + \omega^2 + \epsilon_1\epsilon_2\right) + t_0^4}, \\
\bar{G}_{0,21} &= \frac{t_0 e^{i(\phi_1+\phi_2)}\left(t_0^2 + (\epsilon_1 - \omega)(\omega - \epsilon_2)\right) + \Delta^2 t_0 e^{2i\phi_1}}{e^{i(\phi_1+\phi_2)}\left(\left(\Delta^2 - \omega^2 + \epsilon_1^2\right)\left(\Delta^2 - \omega^2 + \epsilon_2^2\right) - 2t_0^2\left(\omega^2 + \epsilon_1\epsilon_2\right) + t_0^4\right) + \Delta^2 t_0^2 e^{2i\phi_1} + \Delta^2 t_0^2 e^{2i\phi_2}}.
\end{aligned}
\tag{S13}
$$

At this point, we can clearly see that all Green's function components, including the pair amplitudes, exhibit a dependence on the phases, as we indeed expect. Moreover, the intra-superconductor pair amplitudes $F_{11(22)}$ exhibit only even frequency terms. Interestingly, the inter-superconductor pair amplitude $F_{12(21)}$ has a component that is linear in $\omega$ and thus odd in frequency.

As discussed in the main text, we are here interested in the nonlocal Green's function, which involves elements with indices 12 and 21. Consequently, we next restrict our analysis to these inter-superconductor Green's functions. To further simplify the expressions, we consider that $\epsilon_{1,2} = \epsilon$, $\phi_1 = -\phi/2$ and $\phi_2 = \phi/2$. Hence, the inter-superconductor or nonlocal Green's function components are given by

$$\begin{aligned}
F_{12} &= \frac{\Delta t_0 e^{-i\phi/2} \left(-\omega + e^{i\phi}(\omega + \epsilon) + \epsilon\right)}{(\Delta^2 - \omega^2 + \epsilon^2)^2 + 2\Delta^2 t_0^2 \cos(\phi) - 2t_0^2 (\omega^2 + \epsilon^2) + t_0^4}, \\
\bar{F}_{12} &= \frac{\Delta t_0 e^{-i\phi/2} \left(-\omega + e^{i\phi}(\omega + \epsilon) + \epsilon\right)}{(\Delta^2 - \omega^2 + \epsilon^2)^2 + 2\Delta^2 t_0^2 \cos(\phi) - 2t_0^2 (\omega^2 + \epsilon^2) + t_0^4}, \\
F_{21} &= \frac{\Delta t_0 e^{-i\phi/2} \left(\omega + e^{i\phi}(\epsilon - \omega) + \epsilon\right)}{(\Delta^2 - \omega^2 + \epsilon^2)^2 + 2\Delta^2 t_0^2 \cos(\phi) - 2t_0^2 (\omega^2 + \epsilon^2) + t_0^4}, \\
\bar{F}_{21} &= \frac{\Delta t_0 e^{-i\phi/2} \left(\omega + e^{i\phi}(\epsilon - \omega) + \epsilon\right)}{(\Delta^2 - \omega^2 + \epsilon^2)^2 + 2\Delta^2 t_0^2 \cos(\phi) - 2t_0^2 (\omega^2 + \epsilon^2) + t_0^4},
\end{aligned} \quad \text{(S14)}$$

where we see that $F_{12(21)} = \bar{F}_{12(21)}$. For the normal components we get

$$\begin{aligned}
G_{0,12} &= -\frac{t_0 e^{-i\phi} \left(\Delta^2 + e^{i\phi} \left(t_0^2 - (\omega + \epsilon)^2\right)\right)}{(\Delta^2 - \omega^2 + \epsilon^2)^2 - 2t_0^2 (\omega^2 + \epsilon^2) + t_0^4 + 2\Delta^2 t_0^2 \cos(\phi)}, \\
\bar{G}_{0,12} &= \frac{t_0 \left(\Delta^2 e^{i\phi} + t_0^2 - (\epsilon - \omega)^2\right)}{(\Delta^2 - \omega^2 + \epsilon^2)^2 - 2t_0^2 (\omega^2 + \epsilon^2) + t_0^4 + 2\Delta^2 t_0^2 \cos(\phi)}, \\
G_{0,21} &= -\frac{t_0 \left(\Delta^2 e^{i\phi} + t_0^2 - (\omega + \epsilon)^2\right)}{(\Delta^2 - \omega^2 + \epsilon^2)^2 - 2t_0^2 (\omega^2 + \epsilon^2) + t_0^4 + 2\Delta^2 t_0^2 \cos(\phi)}, \\
\bar{G}_{0,21} &= \frac{t_0 e^{-i\phi} \left(\Delta^2 + e^{i\phi}(t_0 - \omega + \epsilon)(t_0 + \omega - \epsilon)\right)}{(\Delta^2 - \omega^2 + \epsilon^2)^2 - 2t_0^2 (\omega^2 + \epsilon^2) + t_0^4 + 2\Delta^2 t_0^2 \cos(\phi)}.
\end{aligned} \quad \text{(S15)}$$

In the main text we also discussed that the superconductor label introduces an additional quantum number that broadens the pair symmetry classes. Thus, the symmetric and antisymmetric combination in the superconductor index can be obtained as $F_{12}^\pm = (F_{12} \pm F_{21})/2$, and similarly for $\bar{F}_{12}^\pm = (\bar{F}_{12} \pm \bar{F}_{21})/2$. We thus find

$$\begin{aligned}
F_{12}^+ &= \frac{2\Delta t_0 \epsilon \cos(\phi/2)}{(\Delta^2 - \omega^2 + \epsilon^2)^2 - 2t_0^2 (\omega^2 + \epsilon^2) + t_0^4 + 2\Delta^2 t_0^2 \cos(\phi)}, \\
F_{12}^- &= \frac{2i\Delta t_0 \omega \sin(\phi/2)}{(\Delta^2 - \omega^2 + \epsilon^2)^2 - 2t_0^2 (\omega^2 + \epsilon^2) + t_0^4 + 2\Delta^2 t_0^2 \cos(\phi)},
\end{aligned} \quad \text{(S16)}$$

which correspond to Eqs. (2) in the main text. We clearly see that $F_-$ is indeed an odd function of the frequency, corresponding to an odd-frequency spin-singlet pair correlation because the parent superconductor is of spin-singlet nature. These expressions clearly show the coexistence of even- and odd-frequency pair amplitudes, which in this case oscillate with a different periodic function in the phase difference $\phi$. Both pair amplitudes are linear in the tunneling parameter $t_0$. While the even-frequency amplitude is linear in $\epsilon$, the odd-frequency component is linear in $\omega$. The linear dependence on $\epsilon$ of the even-frequency part implies that its size can be controlled by the energy of the superconducting quantum dots. In principle, it can be tuned to zero at $\epsilon = 0$, thus leaving odd-frequency pairing as the only finite inter-superconductor pair correlation in Josephson junctions formed by two superconductors. Moreover, at $\epsilon \neq 0$, the even-frequency part $F_{12}^+$ also vanishes at $\phi = \pi$ but the odd-frequency pairing here remains finite. This discussion is presented in the first part of the section on "*Inter-superconductor pair amplitudes in JJs*". Furthermore, we have verified that the vanishing (finite) even-frequency (odd-frequency) pairing at $\phi = \pi$ is an effect that also remains in Josephson junctions with large superconductors, see the last section of this Supplementary Material. Before ending this part, we note that previous works reported the formation of pure odd-$\omega$ pairing in conventional superconductor junctions [1–4]. However, odd-$\omega$ pairing in such studies originated as an intra-superconductor correlation, which points out fundamental differences with the inter-superconductor odd-$\omega$ states studied here.

### Josephson junctions with three superconductors

For Josephson junctions with three superconductors, we proceed as in the previous section. Here, we only list the final results of the pair amplitudes. The nonlocal pair amplitudes at $\epsilon_{1,2,3} = \epsilon$ are given by

$$\begin{aligned}
F_{12}^+ &= \frac{\Delta t_0 \left(e^{i\phi_1} + e^{i\phi_2}\right)\left(\Delta^2\epsilon - t_0^2\epsilon + t_0^3 - t_0\epsilon^2 - \omega^2(t_0+\epsilon) + \epsilon^3\right)}{\left[(\Delta^2 - \omega^2) + (t_0 - \epsilon)^2\right]Q} + P_{12}\,,\\
F_{13}^+ &= \frac{\Delta t_0 \left(e^{i\phi_1} + e^{i\phi_3}\right)\left(\Delta^2\epsilon - t_0^2\epsilon + t_0^3 - t_0\epsilon^2 - \omega^2(t_0+\epsilon) + \epsilon^3\right)}{\left[(\Delta^2 - \omega^2) + (t_0 - \epsilon)^2\right]Q} + P_{13}\,,\\
F_{12}^- &= \frac{\Delta t_0 \omega \left(e^{i\phi_2} - e^{i\phi_1}\right)}{Q}\,,\\
F_{13}^- &= \frac{\Delta t_0 \omega \left(e^{i\phi_3} - e^{i\phi_1}\right)}{Q}\,,
\end{aligned} \tag{S17}$$

with

$$\begin{aligned}
P_{12} &= -\Delta t_0 \frac{-\Delta^2 t_0 e^{i(\phi_1+\phi_2-\phi_3)} + t_0 e^{i\phi_3}[(t_0-\epsilon)^2 - \omega^2]}{\left[(\Delta^2-\omega^2) + (t_0-\epsilon)^2\right]Q}\,,\\
Q &= \left(\Delta^2 - \omega^2 + \epsilon^2\right)^2 - t_0^2\left(\Delta^2 + 5\omega^2 + 3\epsilon^2\right) - 4t_0^3\epsilon + 4t_0^4 + 2t_0\epsilon\left(\Delta^2 - \omega^2 + \epsilon^2\right)\\
&\quad + 2\Delta^2 t_0^2\left[\cos(\phi_1 - \phi_2) + \cos(\phi_1 - \phi_3) + \cos(\phi_2 - \phi_3)\right],
\end{aligned}$$

and $P_{13} = P_{12}(\phi_{2(3)} \to \phi_{3(2)})$.

The previous expressions reduce to simpler forms when expanding around $t_0 = 0$ and keeping only linear terms in $t_0$, namely,

$$\begin{aligned}
F_{12}^+ &\approx \frac{\Delta t_0 \epsilon \left(e^{i\phi_1} + e^{i\phi_2}\right)}{\left(\Delta^2 - \omega^2 + \epsilon^2\right)^2}\,,\\
F_{13}^+ &\approx \frac{\Delta t_0 \epsilon \left(e^{i\phi_1} + e^{i\phi_3}\right)}{\left(\Delta^2 - \omega^2 + \epsilon^2\right)^2}\,,\\
F_{12}^- &\approx \frac{\Delta t_0 \omega \left(e^{i\phi_2} - e^{i\phi_1}\right)}{\left(\Delta^2 - \omega^2 + \epsilon^2\right)^2}\,,\\
F_{13}^- &\approx \frac{\Delta t_0 \omega \left(e^{i\phi_3} - e^{i\phi_1}\right)}{\left(\Delta^2 - \omega^2 + \epsilon^2\right)^2}\,.
\end{aligned} \tag{S18}$$

For four superconductors the expressions get more complicated but, up to first order in $t_0$, we obtain the same expressions as given by Eq. (S18). We have verified that this behavior of the inter-superconductor pair amplitudes between nearest neighbor superconductors remains the same for any number of superconductors within linear order in $t_0$. Hence, we concluded that symmetric even-$\omega$ and antisymmetric odd-$\omega$ pair correlations acquire the phase dependence presented in Eqs. (3) of the main text.

### GREEN'S FUNCTIONS IN JOSEPHSON JUNCTIONS WITH SUPERCONDUCTORS COUPLED VIA A NORMAL REGION

We now obtain the Green's functions for Josephson junctions where superconductors are not coupled directly but, via a normal region instead, see Fig. S1. To model these JJs, we employ the following Hamiltonian

$$H_{\mathrm{nJJ}}^{(2)} = \sum_{j=1}^{n}[\epsilon_j c_{j\sigma}^\dagger c_{j\sigma} + \Delta e^{i\phi_j} c_{j\sigma}^\dagger c_{j\bar{\sigma}}^\dagger] + H_{\mathrm{N}} + H_{\mathrm{T}}^{(2)}\,, \tag{S19}$$

where the first two terms describe the superconductor $S_j$, where $c_{j\sigma}$ ($c_{j\sigma}^\dagger$) annihilates (creates) an electronic state with spin $\sigma$ at site $j$ with onsite energy $\epsilon_j$, phase $\phi_j$, and induced pair potential. Moreover, $H_{\mathrm{N}} = \epsilon_{\mathrm{N}} c_{\mathrm{N}\sigma}^\dagger c_{\mathrm{N}\sigma}$ describes the normal region N, while $H_{\mathrm{T}}^{(2)} = t_0 \sum_{i=1}^{n} c_{i\sigma}^\dagger c_{\mathrm{N}\sigma} + \mathrm{h.c.}$ represents the tunneling Hamiltonian. As in the main text for

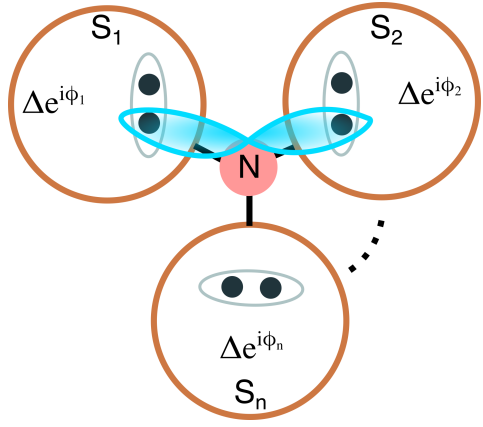

FIG. S1. JJs formed by coupling superconductors $S_i$ with distinct phases $\phi_i$, and same induced pair potential $\Delta$, via a normal region N. In each $S_i$ local pairs are depicted in gray ellipses containing two electrons (black filled circles), referred to as intra superconductor (local) pairs. Due to the tunneling between superconductors, inter-superconductor (nonlocal) pair correlations emerge (cyan) which can be controlled by $\phi_i$.

JJs with superconductors coupled directly, in what follows we drop the spin index for simplicity but keep in mind that the superconductors in Eq. (S19) are spin-singlet. As in previous sections, the pair amplitudes are then obtained from the anomalous component of the Nambu Green's function associated to $H_{2\text{JJ}}^{(2)}$ in Eq. (S19). Specially, from Eqs. S10 and S10 but then using $H_{2\text{JJ}}^{(2)}$ instead of $H_{2\text{JJ}}$.

In this case there are two types of nonlocal pair amplitudes: pair amplitudes between superconductors, also referred before to as inter-superconductor pair amplitudes, and amplitudes between superconductors and the normal region. Below we present the calculations of all of these nonlocal pair amplitudes.

### Josephson junctions with two superconductors

For simplicity, we take $\phi_1 = -\phi/2$, $\phi_2 = \phi/2$, and $\epsilon_{1,2} = \epsilon$ so the nonlocal pair amplitudes are given by

$$
\begin{aligned}
F_{1N}^+ &= \frac{\Delta t_0 e^{-i\phi/2}\left(\epsilon_N\left(\Delta^2 - \omega^2 + \epsilon^2\right) + 2i\epsilon\, t_0^2 \sin(\phi/2)e^{i\phi/2}\right)}{\left(\epsilon_N^2 - \omega^2\right)\left(\Delta^2 - \omega^2 + \epsilon^2\right)^2 + 2t_0^4\left(\Delta^2 - 2\omega^2 + 2\epsilon^2\right) - 4t_0^2\left(\Delta^2 - \omega^2 + \epsilon^2\right)\left(\omega^2 + \epsilon\epsilon_N\right) + 2\Delta^2 t_0^4 \cos(\phi)}, \\
F_{1N}^- &= \frac{\Delta t_0 \omega e^{-i\phi/2}\left(-\Delta^2 + \omega^2 - \epsilon^2 + 2it_0^2 \sin(\phi/2)e^{i\phi/2}\right)}{\left(\epsilon_N^2 - \omega^2\right)\left(\Delta^2 - \omega^2 + \epsilon^2\right)^2 + 2t_0^4\left(\Delta^2 - 2\omega^2 + 2\epsilon^2\right) - 4t_0^2\left(\Delta^2 - \omega^2 + \epsilon^2\right)\left(\omega^2 + \epsilon\epsilon_N\right) + 2\Delta^2 t_0^4 \cos(\phi)}, \\
F_{12}^+ &= \frac{4\Delta\, t_0^2\left(t_0^2 - \omega^2 - \epsilon\epsilon_N\right)\cos(\phi/2)}{\left(\epsilon_N^2 - \omega^2\right)\left(\Delta^2 - \omega^2 + \epsilon^2\right)^2 + 2t_0^4\left(\Delta^2 - 2\omega^2 + 2\epsilon^2\right) - 4t_0^2\left(\Delta^2 - \omega^2 + \epsilon^2\right)\left(\omega^2 + \epsilon\epsilon_N\right) + 2\Delta^2 t_0^4 \cos(\phi)}, \\
F_{12}^- &= -\frac{4i\omega\, \Delta\, t_0^2\left(\epsilon + \epsilon_N\right)\sin(\phi/2)}{\left(\epsilon_N^2 - \omega^2\right)\left(\Delta^2 - \omega^2 + \epsilon^2\right)^2 + 2t_0^4\left(\Delta^2 - 2\omega^2 + 2\epsilon^2\right) - 4t_0^2\left(\Delta^2 - \omega^2 + \epsilon^2\right)\left(\omega^2 + \epsilon\epsilon_N\right) + 2\Delta^2 t_0^4 \cos(\phi)},
\end{aligned}
\tag{S20}
$$

where $\epsilon_N$ here represents the onsite energy in the N region, while $\epsilon$ is the onsite energy of the superconductors. By symmetry, the pair amplitudes between N and superconductor $S_2$, $F_{2N}^\pm$, has the same expression as $F_{1N}^\pm$. We observe that the antisymmetric pair amplitudes acquire a linear dependence on $\omega$ and, therefore, signal the formation of odd-frequency pairing.

We can further simplify previous expressions by expanding up to linear order in $t_0$,

$$F_{1N}^+ \approx -\frac{\Delta\, t_0\, \epsilon_N\, e^{-i\phi/2}}{(\omega^2 - \epsilon_N^2)\,(\Delta^2 - \omega^2 + \epsilon^2)},$$

$$F_{1N}^- \approx \frac{\Delta\, t_0\, \omega\, e^{-i\phi/2}}{(\omega^2 - \epsilon_N^2)\,(\Delta^2 - \omega^2 + \epsilon^2)},$$

$$F_{12}^+ \approx \frac{4\Delta t_0^2\,(\omega^2 + \epsilon\epsilon_N)\cos(\phi/2)}{(\omega^2 - \epsilon_N^2)\,(\Delta^2 - \omega^2 + \epsilon^2)^2},$$

$$F_{12}^- \approx \frac{4i\omega\,\Delta t_0^2(\epsilon + \epsilon_N)\sin(\phi/2)}{(\omega^2 - \epsilon_N^2)\,(\Delta^2 - \omega^2 + \epsilon^2)^2}.$$

(S21)

These pair amplitudes are similar to the ones obtained in the previous section for superconductors coupled directly, but with some interesting differences. First, Eqs. (S21) are proportional to $t_0^2$ while for superconductors coupled directly the pair amplitudes are linear in $t_0$. This effect occurs because coupling via the N region requires two tunneling processes instead of only one for superconductors coupled directly. Second, both inter-superconductor pair amplitudes given by Eqs. (S21) acquire a strong dependence on the onsite energies of the normal and superconducting regions.

### Josephson junctions with three superconductors

The Green's functions become now more complicated, but it is still possible to obtain analytical expressions in some regimes. We are interested in the inter-superconductor pair amplitudes as the superconductor index enables odd-frequency Cooper pairs. The resulting expressions reduce to relatively simple forms for $\epsilon_{1,2,3} = 0$,

$$F_{13}^+ = -\frac{\Delta^3 t_0^4 e^{i(\phi_1 - \phi_2 + \phi_3)} + \Delta t_0^2\left(e^{i\phi_1} + e^{i\phi_3}\right)\left(-\omega^2\left(\Delta^2 + 2t_0^2\right) + \Delta^2 t_0^2 + \omega^4\right) + \Delta t_0^4 \omega^2 e^{i\phi_2}}{(\Delta^2 - \omega^2)\left((\Delta^2 - \omega^2)^2(\omega^2 - \epsilon_N^2) - 3t_0^4(\Delta^2 - 3\omega^2) - 2\Delta^2 t_0^4(\cos(\phi_1 - \phi_2) + \cos(\phi_1 - \phi_3) + \cos(\phi_2 - \phi_3)) + 6t_0^2\omega^2(\Delta^2 - \omega^2)\right)},$$

$$F_{13}^- = \frac{\Delta t_0^2 \omega \epsilon_N \left(e^{i\phi_1} - e^{i\phi_3}\right)}{(\Delta^2 - \omega^2)^2(\epsilon_N^2 - \omega^2) + 3t_0^4(\Delta^2 - 3\omega^2) + 6t_0^2\omega^2(\omega^2 - \Delta^2) + 2\Delta^2 t_0^4(\cos(\phi_1 - \phi_2) + \cos(\phi_1 - \phi_3) + \cos(\phi_2 - \phi_3))},$$

$$F_{12}^+ = -\frac{\Delta t_0^2\left(e^{i\phi_1} + e^{i\phi_2}\right)\left(-\omega^2\left(\Delta^2 + 2t_0^2\right) + \Delta^2 t_0^2 + \omega^4\right) + \Delta^3 t_0^4 e^{i(\phi_1 + \phi_2 - \phi_3)} + \Delta t_0^4 \omega^2 e^{i\phi_3}}{(\Delta^2 - \omega^2)\left((\Delta^2 - \omega^2)^2(\omega^2 - \epsilon_N^2) - 3t_0^4(\Delta^2 - 3\omega^2) - 2\Delta^2 t_0^4(\cos(\phi_1 - \phi_2) + \cos(\phi_1 - \phi_3) + \cos(\phi_2 - \phi_3)) + 6t_0^2\omega^2(\Delta^2 - \omega^2)\right)},$$

$$F_{12}^- = \frac{\Delta t_0^2 \omega \epsilon_N \left(e^{i\phi_1} - e^{i\phi_2}\right)}{(\Delta^2 - \omega^2)^2(\epsilon_N^2 - \omega^2) + 3t_0^4(\Delta^2 - 3\omega^2) + 6t_0^2\omega^2(\omega^2 - \Delta^2) + 2\Delta^2 t_0^4(\cos(\phi_1 - \phi_2) + \cos(\phi_1 - \phi_3) + \cos(\phi_2 - \phi_3))},$$

$$F_{1N}^+ = \frac{\Delta t_0 \epsilon_N e^{i\phi_1}(\Delta - \omega)(\Delta + \omega)}{(\Delta^2 - \omega^2)^2(\epsilon_N^2 - \omega^2) + 3t_0^4(\Delta^2 - 3\omega^2) + 6t_0^2\omega^2(\omega^2 - \Delta^2) + 2\Delta^2 t_0^4(\cos(\phi_1 - \phi_2) + \cos(\phi_1 - \phi_3) + \cos(\phi_2 - \phi_3))},$$

$$F_{1N}^- = \frac{\omega \Delta t_0 \left(e^{i\phi_1}\left(-\Delta^2 - 2t_0^2 + \omega^2\right) + t_0^2\left(e^{i\phi_2} + e^{i\phi_3}\right)\right)}{(\Delta^2 - \omega^2)^2(\epsilon_N^2 - \omega^2) + 3t_0^4(\Delta^2 - 3\omega^2) + 6t_0^2\omega^2(\omega^2 - \Delta^2) + 2\Delta^2 t_0^4(\cos(\phi_1 - \phi_2) + \cos(\phi_1 - \phi_3) + \cos(\phi_2 - \phi_3))}.$$

(S22)

We can obtain the rest of the pair amplitudes, $F_{2(3)N}^\pm$, analogously, but it is enough with the above expressions to see the role of the superconductor index and superconducting phases for generating odd-frequency pairs. An immediate observation in Eqs. (S22) is that the antisymmetric pair amplitudes, $F_{S_1N}^-$ and $F_{S_1S_2}^-$, become odd functions of the frequency $\omega$.

To second order in $t_0$, but at finite onsite energies, the inter-superconductor pair amplitudes are given by

$$F_{12}^+ \approx \frac{\Delta t_0^2\left(e^{i\phi_1} + e^{i\phi_2}\right)\left(\omega^2 + \epsilon\epsilon_N\right)}{(\omega^2 - \epsilon_N^2)\,(\Delta^2 - \omega^2 + \epsilon^2)^2},$$

$$F_{13}^+ \approx \frac{\Delta t_0^2\left(e^{i\phi_1} + e^{i\phi_3}\right)\left(\omega^2 + \epsilon\epsilon_N\right)}{(\omega^2 - \epsilon_N^2)\,(\Delta^2 - \omega^2 + \epsilon^2)^2},$$

$$F_{12}^- \approx \frac{\Delta t_0^2 \omega(\epsilon + \epsilon_N)\left(e^{i\phi_1} - e^{i\phi_2}\right)}{(\epsilon_N^2 - \omega^2)\,(\Delta^2 - \omega^2 + \epsilon^2)^2},$$

$$F_{13}^- \approx \frac{\Delta t_0^2 \omega(\epsilon + \epsilon_N)\left(e^{i\phi_1} - e^{i\phi_3}\right)}{(\epsilon_N^2 - \omega^2)\,(\Delta^2 - \omega^2 + \epsilon^2)^2},$$

(S23)

which reveal a similar dependence on the phases as when the superconductors are coupled directly, see previous section. Due to the normal region, however, the inter-superconductor pair amplitudes require at least two tunneling processes, instead of the single process needed in Josephson junctions with superconductors coupled directly. In fact, we have verified that for any number of superconductors the inter-superconductor pair amplitudes between nearest neighbors follow the same expressions as those given by Eq. (S23). Expanding the pair amplitudes around $t_0 = 0$ and keeping terms up to second order in $t_0$, we obtain

$$
\begin{aligned}
F^+_{j,j+1}(\omega) &\approx \frac{\Delta t_0^2(\omega^2 - \epsilon\epsilon_{\rm N})\left(e^{i\phi_{j+1}} + e^{i\phi_j}\right)}{(\omega^2 - \epsilon_{\rm N}^2)(\Delta^2 - \omega^2 + \epsilon^2)^2}, \\
F^-_{j,j+1}(\omega) &\approx \frac{\Delta \omega t_0^2(\epsilon + \epsilon_{\rm N})\left(e^{i\phi_{j+1}} - e^{i\phi_j}\right)}{(\omega^2 - \epsilon_{\rm N}^2)(\Delta^2 - \omega^2 + \epsilon^2)^2},
\end{aligned}
\quad (S24)
$$

for the symmetric and antisymmetric pair amplitudes, respectively. Here, $\epsilon_{\rm N}$ ($\epsilon$) represent onsite energies of the N (S) regions. The pair amplitudes in Eqs. (S24) exhibit some similarities but also differences in relation to the pair amplitudes in Josephson junctions with superconductors coupled directly [Eqs. (3) of the main text]. Among the similarities, we find that both develop the same dependence on the superconducting phases, namely, the symmetric even-$\omega$ (antisymmetric odd-$\omega$) pair amplitudes are proportional to $(e^{i\phi_{j+1}} \pm e^{i\phi_j})$: the even-$\omega$ term vanishes when the phase difference between the involved superconductors is $\pi$ but the odd-$\omega$ part remains finite, see Eqs. (S24).

On the differences, we find that both pair amplitudes strongly depend on the onsite energies $\epsilon_{\rm N}$ ($\epsilon$). On the one hand, at $\epsilon = 0$ and $\epsilon_{\rm N} \neq 0$, the pair amplitudes are given by $F^+_{j,j+1} \sim \omega^2(e^{i\phi_{j+1}} + e^{i\phi_j})$ and $F^-_{j,j+1} \sim \omega\epsilon_{\rm N}(e^{i\phi_{j+1}} - e^{i\phi_j})$, with a ratio given by $\sim (\omega/\epsilon_{\rm N})(e^{i\phi_{j+1}} + e^{i\phi_j})/(e^{i\phi_{j+1}} - e^{i\phi_j})$. On the other hand, at $\epsilon_{\rm N} = 0$ and $\epsilon \neq 0$, we obtain $F^+_{j,j+1} \sim (e^{i\phi_{j+1}} + e^{i\phi_j})$ and $F^-_{j,j+1} \sim (\epsilon/\omega)(e^{i\phi_{j+1}} - e^{i\phi_j})$, with a ratio given by $\sim (\epsilon/\omega)(e^{i\phi_{j+1}} + e^{i\phi_j})/(e^{i\phi_{j+1}} - e^{i\phi_j})$. Thus, we can see that the formation of inter-superconductor pair amplitudes, specially their ratio, in Josephson junctions coupled via normal regions can be controlled by the superconducting phases and also by the onsite energies.

Finally, we would like to mention that modelling superconductors coupled to a quantum dot, as we consider here, in principle requires to take Coulomb interactions into account. These interactions are expected due to the reduced dimensionality of quantum dots and are often captured in terms of a charging energy that, together with the coupling to the superconductor and order parameter, has been shown to be crucial for the induced superconducting state in the quantum dot [5]. Commonly, in the regime where Kondo and exchange interactions between superconductors can be neglected [6], the local Coulomb interactions are treated by a Hartree-Fock approximation which has shown to produce a renormalization of the onsite energies and the coupling between the quantum dot and the superconductors [7–10]. While this approach has been useful to qualitatively explain certain experiments [10], it is fair to say that more refined approaches [11, 12] are required to quantitatively describe the emergent superconducting states in quantum dots coupled to superconductors. In this regard, the role of the charging energy on the formation of odd-$\omega$ pairs and how this interplay can be controlled by means of the superconducting phases represents an interesting open problem that could open a new avenue for realizing odd-$\omega$ superconductivity.

## CROSSED ANDREEV REFLECTIONS IN JOSEPHSON JUNCTIONS WITH TWO SUPERCONDUCTORS

We now provide further details on how to obtain the crossed Andreev reflections (CAR) in a Josephson junction with two superconductors. We also provide details of the electron tunneling (ET) processes as both scattering mechanisms finally determine the nonlocal conductance. The ET processes are obtained as $T_{\rm ET} = T_{ee} + T_{hh}$ with

$$
\begin{aligned}
T_{ee}(\omega) &= \Gamma_{\rm L}^e \Gamma_{\rm R}^e |G^r_{0,12}(\omega)|^2, \\
T_{hh}(\omega) &= \Gamma_{\rm L}^h \Gamma_{\rm R}^h(\omega) |\bar{G}^r_{0,12}|^2,
\end{aligned}
\quad (S25)
$$

while the CAR processes are $T_{\rm CAR} = T_{eh} + T_{he}$ with

$$
\begin{aligned}
T_{eh}(\omega) &= \Gamma_{\rm L}^e \Gamma_{\rm R}^h |F^r_{12}(\omega)|^2, \\
T_{he}(\omega) &= \Gamma_{\rm L}^h \Gamma_{\rm R}^e |\bar{F}^r_{12}(\omega)|^2,
\end{aligned}
\quad (S26)
$$

where $G^r_{0,12}$ ($\bar{G}^r_{0,12}$) and $F^r_{12}$ ($\bar{F}^r_{12}$) correspond to the normal and anomalous components of the retarded Green's function $G^r$, which is obtained by including the effect of the leads as retarded self-energies $\Sigma^r_j$,

$$
G^r(\omega) = (\omega - H_{\rm 2JJ} - \Sigma^r_{\rm L} - \Sigma^r_{\rm R})^{-1}. \quad (S27)
$$

Moreover, in the wide-band limit, the self-energies are given by $\Sigma_j = -i\Gamma_j/2$, where $\Gamma_j = \pi|\tau|^2\rho_j$ characterizes the couplings of the leads $j = L, R$ to superconductor $S_j$, being $\rho_j$ the surface density of states and $\tau$ the hopping between leads and superconductors. In Eqs. (S25) and (S26), $\Gamma_j^{e(h)}$ correspond to the electron (hole) component of the coupling matrix $\Gamma_j$ for normal lead $j = L, R$, see also Fig. 1 in the main text.

The nonlocal conductance is determined by CAR and ET processes as

$$G_{\rm LR} = dI_{\rm L}/dV_{\rm R} = (2e^2/h)[T_{\rm CAR} - T_{\rm EC}]. \tag{S28}$$

which reveals that regimes with $T_{\rm CAR} \gg T_{\rm EC}$ provide direct information of the emerging superconducting pair corrections, see also Eq. (S26). Before going further we also stress that, for sufficiently small $\Gamma_i$, the properties of the Josephson junction are not altered and can be probed by a transport experiment described by Eq. (S28), where the involved processes are characterized by normal and anomalous components of the nonlocal (normal and anomalous) Green's functions as described given by Eqs. (S25) and (S26). Below we calculate the required Green's functions.

We obtain,

$$G_{0,12}^r = \frac{t_0\left(\frac{i\Gamma_{\rm L}}{2} + \omega + \epsilon\right)\left(\frac{i\Gamma_{\rm R}}{2} + \omega + \epsilon\right) - t_0\left(t_0^2 + \Delta^2 e^{i(\phi_1 - \phi_2)}\right)}{\frac{1}{16}(4(\Delta^2 + \epsilon^2) + (\Gamma_{\rm L} - 2i\omega)^2)(4(\Delta^2 + \epsilon^2) + (\Gamma_{\rm R} - 2i\omega)^2) + \frac{1}{2}t_0^2(-4\epsilon^2 + (\Gamma_{\rm L} - 2i\omega)(\Gamma_{\rm R} - 2i\omega)) + 2\Delta^2 t_0^2 \cos(\phi_1 - \phi_2) + t_0^4},$$

$$\bar{G}_{0,12}^r = \frac{t_0\left(\frac{1}{4}(\Gamma_{\rm L} + 2i(\epsilon - \omega))(\Gamma_{\rm R} + 2i(\epsilon - \omega)) + \Delta^2 e^{-i(\phi_1 - \phi_2)} + t_0^2\right)}{\frac{1}{16}(4(\Delta^2 + \epsilon^2) + (\Gamma_{\rm L} - 2i\omega)^2)(4(\Delta^2 + \epsilon^2) + (\Gamma_{\rm R} - 2i\omega)^2) + \frac{1}{2}t_0^2(-4\epsilon^2 + (\Gamma_{\rm L} - 2i\omega)(\Gamma_{\rm R} - 2i\omega)) + 2\Delta^2 t_0^2 \cos(\phi_1 - \phi_2) + t_0^4},$$
$$\tag{S29}$$

for the normal components and,

$$F_{12}^{r,+} = \frac{8\Delta t_0 \epsilon \left(e^{i\phi_1} + e^{i\phi_2}\right)}{C_3 + i\omega C_4},$$
$$F_{21}^{r,-} = \frac{2\Delta t_0 \omega \left(e^{i\phi_2} - e^{i\phi_1}\right) + i\Delta t_0 \left(\Gamma_{\rm L} e^{i\phi_2} - \Gamma_{\rm R} e^{i\phi_1}\right)}{C_3 + i\omega C_4}, \tag{S30}$$

for the anomalous symmetric and antisymmetric components ($F^{\pm} = (F_{12} \pm F_{21})/2$), where

$$C_3 = \left[2\omega^2 - 2\left(\frac{\Gamma_{\rm L}^2}{4} + \Gamma_{\rm L}\Gamma_{\rm R} + \frac{\Gamma_{\rm R}^2}{4} + 2\left(\Delta^2 + \epsilon^2\right) + 2t_0^2\right)\right]\omega^2 + \frac{1}{8}Ae^{-i(\phi_1+\phi_2)},$$
$$A = e^{i(\phi_1+\phi_2)}\left(\left(\Gamma_{\rm L}^2 + 4\left(\Delta^2 + \epsilon^2\right)\right)\left(\Gamma_{\rm R}^2 + 4\left(\Delta^2 + \epsilon^2\right)\right) + 8t_0^2\left(\Gamma_{\rm L}\Gamma_{\rm R} - 4\epsilon^2\right) + 16t_0^4\right) + 16\Delta^2 t_0^2 e^{2i\phi_1} + 16\Delta^2 t_0^2 e^{2i\phi_2},$$
$$C_4 = 2\left[\omega^2 - \left(\frac{\Gamma_{\rm L}\Gamma_{\rm R}}{4} + \Delta^2 + t_0^2 + \epsilon^2\right)\right](\Gamma_{\rm L} + \Gamma_{\rm R}).$$
$$\tag{S31}$$

We also find that,

$$\bar{F}_{12}^{r,+} = e^{-i(\phi_1+\phi_2)} F_{12}^{r,+}(\Gamma_{\rm L} \to \Gamma_{\rm R}, \Gamma_{\rm R} \to \Gamma_{\rm L}),$$
$$\bar{F}_{12}^{r,-} = e^{-i(\phi_1+\phi_2)} F_{12}^{r,-}(\Gamma_{\rm L} \to \Gamma_{\rm R}, \Gamma_{\rm R} \to \Gamma_{\rm L}). \tag{S32}$$

### Equal couplings to the leads

Considering that the couplings to the leads are symmetric, $\Gamma_{L,R} = \Gamma$, $G^r(\omega)$ can be simply obtained by replacing $\omega \to \omega + i0^+ + i\Gamma/2$ in the Green's function $G_{\rm 2JJ}$ given by Eqs. (S10), (S12), and (S13). In this case, the expressions

given above reduce to

$$\begin{aligned}
G_{0,12}^r &= -\frac{4t_0\left(4\Delta^2 + e^{i\phi}\left(4t_0^2 + (\Gamma - 2i(\omega+\epsilon))^2\right)\right)}{e^{i\phi}\left((4(\Delta^2+\epsilon^2) + (\Gamma - 2i\omega)^2)^2 + 8t_0^2(-4\epsilon^2 + (\Gamma - 2i\omega)^2) + 16t_0^4\right) + 16\Delta^2 t_0^2 e^{2i\phi} + 16\Delta^2 t_0^2}, \\
\bar{G}_{0,12}^r &= \frac{4t_0 e^{i\phi}\left((\Gamma + 2i(\epsilon-\omega))^2 + 4\Delta^2 e^{i\phi} + 4t_0^2\right)}{e^{i\phi}\left((4(\Delta^2+\epsilon^2) + (\Gamma - 2i\omega)^2)^2 + 8t_0^2(-4\epsilon^2 + (\Gamma - 2i\omega)^2) + 16t_0^4\right) + 16\Delta^2 t_0^2 e^{2i\phi} + 16\Delta^2 t_0^2}, \\
F_{12}^{r,+} &= \frac{16\Delta t_0 \epsilon e^{\frac{i\phi}{2}}\left(1 + e^{i\phi}\right)}{e^{i\phi}\left((4(\Delta^2+\epsilon^2) + (\Gamma - 2i\omega)^2)^2 + 8t_0^2(-4\epsilon^2 + (\Gamma - 2i\omega)^2) + 16t_0^4\right) + 16\Delta^2 t_0^2 e^{2i\phi} + 16\Delta^2 t_0^2}, \\
F_{12}^{r,-} &= \frac{8\Delta t_0 e^{\frac{i\phi}{2}}\left(-1 + e^{i\phi}\right)(2\omega + i\Gamma)}{e^{i\phi}\left((4(\Delta^2+\epsilon^2) + (\Gamma - 2i\omega)^2)^2 + 8t_0^2(-4\epsilon^2 + (\Gamma - 2i\omega)^2) + 16t_0^4\right) + 16\Delta^2 t_0^2 e^{2i\phi} + 16\Delta^2 t_0^2}.
\end{aligned} \tag{S33}$$

Moreover, at $\epsilon = 0$, the previous equations reduce to

$$\begin{aligned}
G_{0,12}^r &= -\frac{4t_0\left(4\Delta^2 + e^{i\phi}\left(4t_0^2 + (\Gamma - 2i\omega)^2\right)\right)}{e^{i\phi}\left((4\Delta^2 + (\Gamma - 2i\omega)^2)^2 + 8t_0^2(\Gamma - 2i\omega)^2 + 16t_0^4\right) + 16\Delta^2 t_0^2 e^{2i\phi} + 16\Delta^2 t_0^2}, \\
\bar{G}_{0,12}^r &= \frac{4t_0 e^{i\phi}\left((\Gamma - 2i\omega)^2 + 4\Delta^2 e^{i\phi} + 4t_0^2\right)}{e^{i\phi}\left((4\Delta^2 + (\Gamma - 2i\omega)^2)^2 + 8t_0^2(\Gamma - 2i\omega)^2 + 16t_0^4\right) + 16\Delta^2 t_0^2 e^{2i\phi} + 16\Delta^2 t_0^2}, \\
F_{12}^{r,+} &= 0, \\
F_{12}^{r,-} &= \frac{8\Delta t_0 e^{\frac{i\phi}{2}}\left(-1 + e^{i\phi}\right)(2\omega + i\Gamma)}{e^{i\phi}\left((4\Delta^2 + (\Gamma - 2i\omega)^2)^2 + 8t_0^2(\Gamma - 2i\omega)^2 + 16t_0^4\right) + 16\Delta^2 t_0^2 e^{2i\phi} + 16\Delta^2 t_0^2},
\end{aligned} \tag{S34}$$

so that the symmetric component vanishes leaving only the antisymmetric part of the pair amplitudes present. This antisymmetric part is the odd-frequency pairing.

We can also see that, for $\Delta = t$ and $\phi = \pi$, the expressions above become

$$\begin{aligned}
G_{0,12}^r &= -\frac{4t_0(\Gamma - 2i\omega)^2}{(\Gamma - 2i\omega)^2\left[16t_0^2 + (\Gamma - 2i\omega)^2\right]}, \\
\bar{G}_{0,12}^r &= \frac{4t_0(\Gamma - 2i\omega)^2}{(\Gamma - 2i\omega)^2\left[16t_0^2 + (\Gamma - 2i\omega)^2\right]}, \\
F_{12}^{r,+} &= 0, \\
F_{12}^{r,-} &= -\frac{16(\Gamma - 2i\omega)t_0^2}{(\Gamma - 2i\omega)^2\left(16t_0^2 + (\Gamma - 2i\omega)^2\right)},
\end{aligned} \tag{S35}$$

which shows that, for $\phi = \pi$ and $\Delta = t_0$, the contribution of the pair amplitude is larger than $G_{0,12}^r$. Interestingly, this regime corresponds to the closing of the gap in the Andreev bound states of the junction with two superconductors. Eqs. (S34) and (S35) are discussed in Eqs. (6) and also in section on "*CAR detection of odd-$\omega$ pairing*" of the main text where we denote $G_{0,12}^r$ and $\bar{G}_{0,12}^r$ as $g_{12}^r$ and $\bar{g}_{12}^r$, respectively.

Before closing this part, we point out that in JJs with more than two superconductors the analysis gets more complicated but it is still possible to obtain nonlocal transport regimes with dominant odd-$\omega$ pairing. For this purpose, it is important to consider weak tunneling between superconductors, such that the even-$\omega$ pairing vanishes either at $\epsilon = 0$ or when the phase difference between involved superconductors is $\pi$, as revealed by Eqs. (3) of the main text. Consequently, $T_{eh}$ involves only contributions from odd-$\omega$ pairing. Then, by following a similar analysis as for JJs with two superconductors, it is possible to show that $T_{ee(hh)} \ll T_{eh(he)}$ at low-frequencies, provided that the phase difference between the involved superconductors is $\pi$. Therefore, the finite value of nonlocal conductance at low frequencies can still be considered a signature of odd-$\omega$ pairing in JJs with more than two superconductors. This thus supports the discussion made in section on "*CAR detection of odd-$\omega$ pairing*" of the main text.

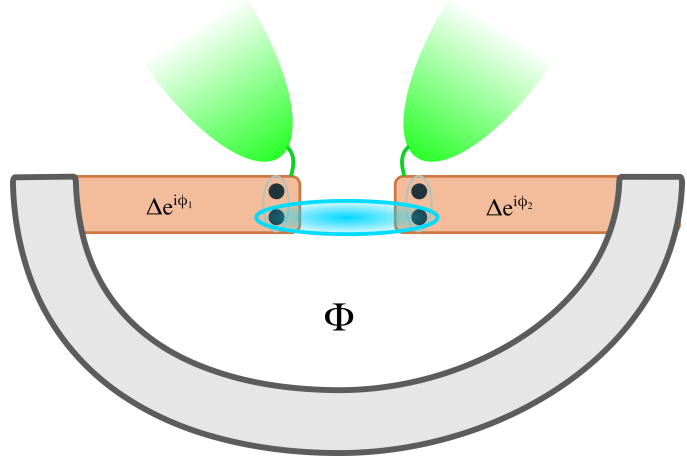

FIG. S2. Sketch of the Josephson junction based on two finite size superconductors (orange) subjected to phases $\phi_{1,2}$ and spin-singlet s-wave pair pontentials $\Delta$. The superconductors are placed in a loop SQUID circuit (gray) such that their phase difference $\phi = \phi_2 - \phi_1$ is controlled by an external flux $\Phi$. As discussed in this SM and main text, tunneling between superconductors induces inter-superconductor (nonlocal) pair correlations (cyan) which can be controlled by $\phi$ and measured in CAR processes by attaching normal leads (green).

## NONLOCAL EVEN- AND ODD-$\omega$ PAIR AMPLITUDES IN JOSEPHSON JUNCTIONS WITH FINITE LENGTH SUPERCONDUCTORS

To further support our findings on the dominant behaviour of odd-$\omega$ pair amplitudes, here we consider Josephson junctions formed by superconductors with finite size and realistic parameters as sketched in Fig. S2. For this purpose, we consider conventional spin-singlet s-wave superconductors described by a one-dimensional tight-binding lattice given by

$$H_\alpha = (2t - \mu) \sum_{\sigma n} c^\dagger_{\sigma n \alpha} c_{\sigma n \alpha} - \sum_{\sigma \langle n,n' \rangle} t\, c^\dagger_{\sigma n' \alpha} c_{\sigma n \alpha} + \sum_{\sigma n} \Delta\, e^{i\phi_\alpha}\, c^\dagger_{\sigma n \alpha} c^\dagger_{\bar\sigma n \alpha} + \text{H.\,c.}\,, \tag{S36}$$

where $c^\dagger_{\sigma n \alpha}$ creates a fermionic state with spin $\sigma$ at site $n$ in superconductor $\alpha$. The site index runs from $n = 1_\text{S}$ till $n = N_\text{S}$, with $N_\text{S}$ being the last site of the superconductor $\alpha$ whose length is then $L_\text{S} = N_\text{S} a_0$ where $a_0$ is the lattice spacing. Moreover, $t = \hbar^2/(2m^* a_0^2)$ is the hopping with $m^*$ being the effective mass, $\mu$ is the chemical potential defining the filling of the system, $\langle n,n' \rangle$ denotes nearest-neighbor sites, and $\Delta$ represents the pair potential and $\phi_\alpha$ its phase. To show the dominant behaviour of odd-$\omega$ pairing, we consider a Josephson junction formed by two superconductors modelled by Eq. (S36) coupled directly. These two superconductors are then refereed to as left (L) and right (R) superconductors. In this case, the tunneling Hamiltonian is then given by

$$H_\text{T} = t \sum_\sigma c^\dagger_{\sigma N_\text{S} \text{L}} c_{\sigma 1_\text{S} \text{R}} + \text{H.c.}, \tag{S37}$$

where $c^\dagger_{\sigma N_\text{S} \text{L}}$ creates a fermionic state in the last site of left superconductor while $c_{\sigma 1_\text{S} \text{R}}$ annihilates a fermionic state in the first site of the right superconductor. The tunneling Hamiltonian $H_\text{T}$ does not involve spin-flip processes and corresponds to the same tunneling model given by Eq. (S5) and the one given in Eq. (1) of the main text. Then, a Josephson junction with two superconductors is modelled by the Nambu Hamiltonian $\bar{H}_{2JJ}$ similar to the one given by Eq. (S6) but now with the Hamiltonians for the superconductors given by Eqs. (S36) and for the tunneling term by Eq. (S37). This way we are able to describe a Josephson junction with finite superconductors without relying on single site superconductors.

We now numerically obtain the Green's function $\bar{G}$ associated to the Josephson junction with two finite superconductors. We carry out these calculations for realistic parameters taking a finite phase difference $\phi_R - \phi_L = \phi$ and $t = 25\,\text{meV}$, $a_0 = 10\,\text{nm}$, $\Delta = 0.5\,\text{meV}$, $\mu = 0.5\,\text{meV}$ and also considering realistic lengths of the superconductors ranging between $L_\text{S} = 400\,\text{nm}$ to $L_\text{S} = 2000\,\text{nm}$. We do this by using the equation of motion $[\omega - \bar{H}_{2JJ}]\bar{G} = 1$, where $\bar{H}_{2JJ}$ is the Nambu Hamiltonian describing the Josephson junction with finite superconductors. Once we obtain $\bar{G}$, we

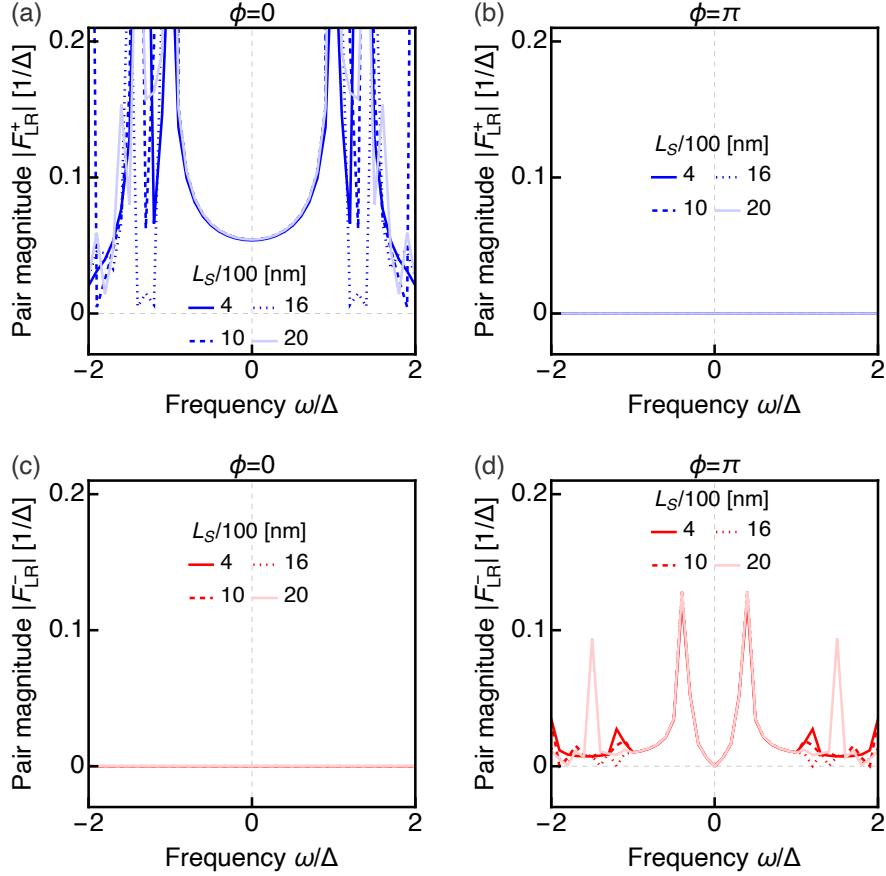

FIG. S3. (a,b) Symmetric even-$\omega$ ($F_{\text{LR}}^{+}$) and (c,d) antisymmetric odd-$\omega$ ($F_{\text{LR}}^{-}$) nonlocal pair amplitudes in a Josephson junction with two finite length superconductors coupled directly as a function of frequency $\omega$ at $\phi=0$ and $\phi=\pi$. Different curves in each panel correspond to different lengths of the superconductors, where solid blue, dashed blue, dotted blue, and light blue correspond to $L_{\text{S}}=400$nm, $L_{\text{S}}=1000$nm, $L_{\text{S}}=1600$nm, and $L_{\text{S}}=2000$nm, respectively. Parameters: $\Delta=0.5$meV, $\mu=0.5$meV.

extract the pair amplitudes between the last site of the left superconductor and the first site of the right superconductor and label them just by $L$ and $R$ indices denoting that they represent pair correlations between left (L) and right (R) superconductors. These nonlocal pair amplitudes $F_{LR}$ are then decomposed into their symmetric and antisymmetric components under the exchange of L and R, which are then denoted by $F_{\text{LR}}^{+}$ and $F_{\text{LR}}^{-}$, respectively. Moreover, we note that since there are no spin-mixing fields, these nonlocal pair amplitudes have spin-singlet symmetry which implies that $F_{\text{LR}}^{\pm}$ corresponds to even-$\omega$ (odd-$\omega$), spin-singlet, even (odd) in superconductor indices. Interestingly, these nonlocal pair amplitudes correspond to the even- and odd-$\omega$ nonlocal pair amplitudes discussed in the main part of our manuscript. In Fig. S3(a-d) and Fig. S4(a,b,d,e) we present the magnitude of these pair amplitudes as a function of frequency $\omega$ and phase difference $\phi$ for distinct lengths of the superconductors. In Fig. S4(c,f) we also show the ratio between even- and odd-$\omega$ pair amplitudes for several realistic lengths of the superconductors.

At zero phase $\phi=0$ only the even-$\omega$ component is finite for frequencies within the gap (and also outside) [Figs. S3(a,c)], while at $\phi=\pi$ it is the odd-$\omega$ pair amplitude the only pair amplitude that remains finite [Figs. S3(b,d)]. Of course that at zero frequency the odd-$\omega$ pairing vanishes as expected for any odd-$\omega$ function. Having odd-$\omega$ pairing as the only type of nonlocal superconducting pairing at $\phi=\pi$ remains even when the length of the superconductors $L_{\text{S}}$ increase, supporting the discovery reported in the first part of our manuscript, specially Fig. 2. The intriguing and interesting behaviour of these pair correlations can be further seen in Fig. S4, where we clearly observe that the even-$\omega$ part completely vanishes at $\phi=\pi$, while the odd-$\omega$ pairing becomes the only finite nonlocal pair amplitude, seen by comparing Fig. S4(a,d) with Fig. S4(b,e). Furthermore, by inspecting the ratio between even- and odd-$\omega$ pair correlations, we note that such ratio vanishes at $\phi=\pi$ due to the vanishing of even-$\omega$ pairing, see Fig. S4(c,f). As a result, we conclude that odd-$\omega$ pairing becomes the only type of nonlocal superconducting pairing at $\phi=\pi$ in Josephson junctions with realistic superconductors. These results are in line with our findings presented in the

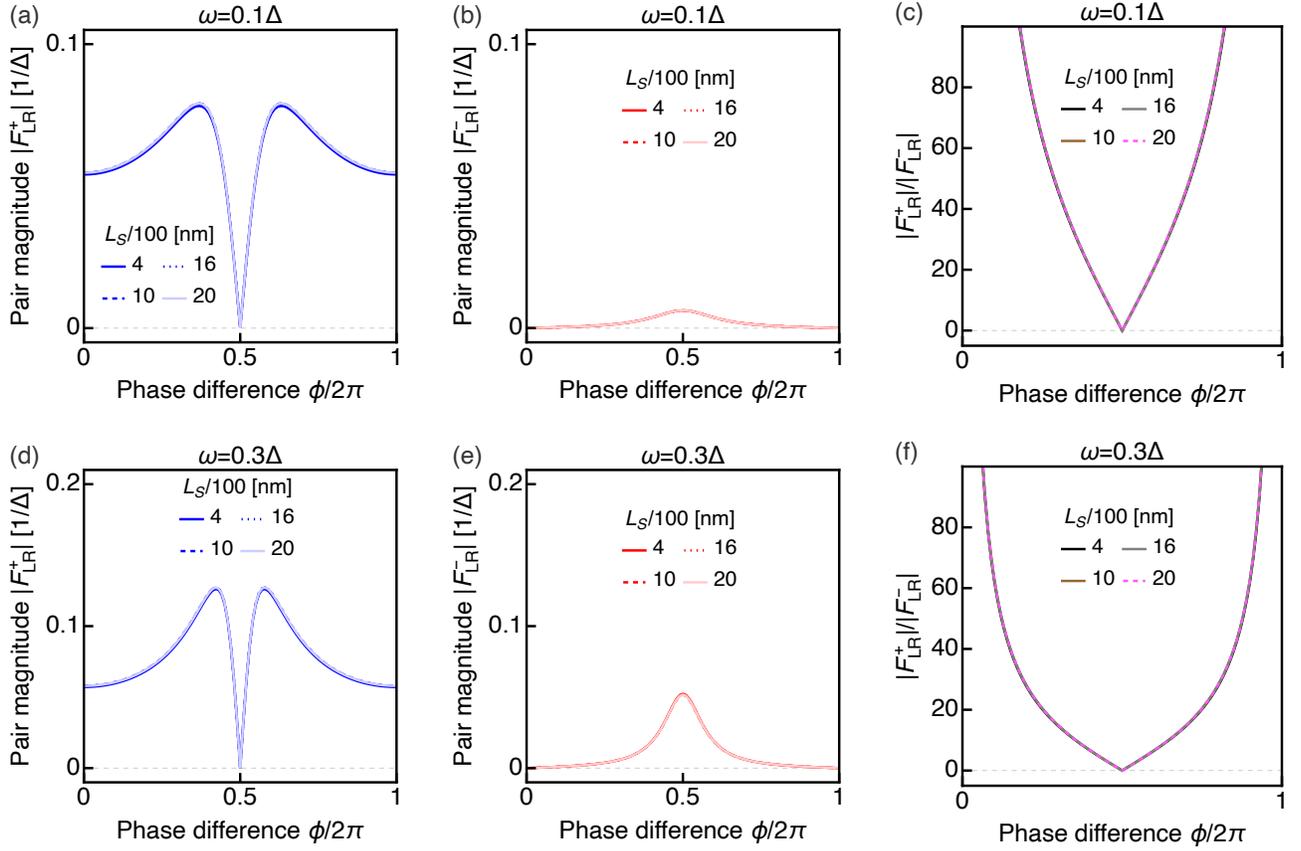

FIG. S4. (a,d) Symmetric even-$\omega$ ($F_{\rm LR}^{+}$) and (b,e) antisymmetric odd-$\omega$ ($F_{\rm LR}^{-}$) nonlocal pair amplitudes in a Josephson junction with two finite length superconductors coupled directly as a function of the phase difference $\phi$ at $\omega = 0.1\Delta$ and $\omega = 0.3\Delta$. In (c,d) the ratio $|F_{\rm LR}^{+}|/|F_{\rm LR}^{-}|$ is presented for each case of (a,b) and (d,e). Different curves in each panel correspond to different lengths of the superconductors, see legends. Parameters: $\Delta = 0.5$meV, $\mu = 0.5$meV.

first part of the main text where, however, Josephson junctions are modelled by single site superconductors. The agreement between the results presented in this section and those shown in the main text clearly demonstrate that our findings about odd-$\omega$ pairing being the only type of nonlocal superconducting pairing at $\phi = \pi$ are robust and very likely to appear even in realistic Josephson junctions. The reason for this agreement is because our simple model in the main text Eq. (1) already captures the tunnelling processes that permits us to explore Josephson transport in multi-superconductor Josephson junctions. As a result, having odd-$\omega$ pairing as the only type of nonlocal pairing implies that it is the main effect for enabling CAR at $\phi = \pi$ exactly in the same way as discussed in the section on "*CAR detection of odd-$\omega$ pairing*" of the main text.


\end{document}